\documentclass[prb,twocolumn,showpacs,floatfix,citesort]{revtex4}
\usepackage{makeidx}
\usepackage{graphicx}
\usepackage{amsmath}
\usepackage{amsfonts}
\usepackage{amssymb}
\usepackage{epsfig}
\usepackage{times}
\usepackage{psfrag}

\begin{document}

\title{The Spin Density Matrix II: Application to a system of two quantum
dots}
\author{Sharif D. Kunikeev}
\affiliation{Department of Chemistry, University of Southern California, Los Angeles, CA
90089}
\author{Daniel A. Lidar}
\affiliation{Departments of Chemistry, Electrical Engineering, and Physics, University of
Southern California, Los Angeles, CA 90089}

\begin{abstract}
This work is a sequel to our work \textquotedblleft The Spin Density Matrix
I: General Theory and Exact Master Equations\textquotedblright\ (eprint
cond-mat/0708.0644). Here we compare \emph{pure}- and \emph{pseudo}-spin
dynamics using as an example a system of two quantum dots, a pair of
localized conduction-band electrons in an $n$-doped GaAs semiconductor. 
\emph{Pure}-spin dynamics is obtained by tracing out the orbital degrees of
freedom, whereas \emph{pseudo}-spin dynamics retains (as is conventional) an
implicit coordinate dependence. We show that magnetic field inhomogeneity
and spin-orbit interaction result in a non-unitary evolution in \emph{pure}%
-spin dynamics, whereas these interactions contribute to the effective \emph{%
pseudo}-spin Hamiltonian via terms that are asymmetric in spin permutations, in
particular, the Dzyaloshinskii-Moriya (DM) spin-orbit interaction. We
numerically investigate the non-unitary effects in the dynamics of the
triplet states population, purity, and Lamb energy shift, as a function of
interdot distance and magnetic field difference $\Delta \vec{B}$. The
spin-orbit interaction is found to produce effects of roughly four orders of
magnitude smaller than those due to $\Delta \vec{B}$ in the \emph{pure}-spin
model. We estimate the spin-orbit interaction magnitude in the
DM-interaction term. Our estimate gives a smaller value than that recently
obtained by Kavokin [Phys. Rev. B \textbf{64}, 075305 (2001)], who did not
include double occupancy effects. We show that a necessary and sufficient
condition for obtaining a universal set of quantum logic gates,
involving only two spins, in both \emph{%
pure}- and \emph{pseudo}-spin models is that the magnetic field
inhomogeneity $\Delta \vec{B}$ and the Heisenberg interaction are both
non-vanishing. We also briefly analyze \emph{pure}-spin dynamics in
the electron on liquid helium system recently proposed by Lyon [Phys. Rev. A 
\textbf{74}, 052338 (2006)].
\end{abstract}

\maketitle

\section{Introduction}

The spin degree of freedom of a localized particle, e.g., an electron or
nucleus, is a popular carrier of quantum information. It serves as a qubit
which can be manipulated in order to accomplish a computational
task. The spin
of electrons localized in quantum dots (QDs) or by donor atoms has been the
subject of extensive recent studies \cite%
{Loss:98,Kane:98,Vrijen:00,HuSarma01,Schliemann01,HuSarma02,Koiller02,Kaplan04,Scarola,He05,Hu,KSS97,KA98,GAP98,AS02}%
.

Consider two electrons trapped in two sites $A$ and $B$, e.g., two QDs \
each containing one electron. The two-electron system is fully described by
the total wavefunction $\left\vert \Psi _{\mathrm{tot}}\right\rangle $,
which depends on the electrons' coordinates $\vec{r}$ \ and spin variables $%
\sigma $. The two-electron spin-density matrix, obtained by tracing out the
orbital degrees of freedom, $\rho ={\rm Tr}_{\vec{r}}\,\left\vert \Psi _{%
\mathrm{tot}}\right\rangle \left\langle \Psi _{\mathrm{tot}}\right\vert $,
fully describes the spin dynamics as long as one cannot or does not wish to
apply measurements that can separate or localize electrons spatially; the
only observable is then the electron spin, $s^{\alpha }=\frac{1}{2}\sigma
_{\alpha }$, where $\sigma _{\alpha }$ are the Pauli spin one-half matrices
with $\alpha =x,y,z$. Since the spin system is not closed -- there is a
coupling to the electrons' spatial degrees of freedom -- we observe open system
effects, i.e., the spin dynamics becomes in general non-unitary. We refer to
this dynamics as \emph{pure}-spin dynamics.

In contrast, \emph{pseudo}-spin dynamics is the standard case where the
electron spin observable is not free from coordinate dependence but includes
information about the electron's localization orbital. In the \emph{pseudo}%
-spin case one defines the electron spin operator as a bilinear combination
of electron annihilation and creation Fermi operators, $c_{As}$, $%
c_{As}^{\dagger }$, in a localized orbital $\phi _{A}$ ($s$ is a spin index, 
$A\ $is the QD\ index): $s_{A}^{\alpha }=\frac{1}{2}\sum_{ss^{\prime
}=1}^{2}c_{As}^{\dagger }\left( \sigma _{\alpha }\right) _{ss^{\prime
}}c_{As^{\prime }}$, $\alpha =x,y,z$. Then the operators $\{s_{A}^{\alpha
}\}_{\alpha }$ obey the usual su$(2)$ commutation rules.

This paper is the sequel to our work Ref.~\onlinecite{KLI07} (henceforth
\textquotedblleft part I\textquotedblright ), where we derived an
operator-sum representation (OSR) as well as a master equation in the
Lindblad and time-convolutionless (TCL) forms for the spin-density matrix of
a two-electron system. In this sequel we focus on a detailed comparison of 
\emph{pure} and \emph{pseudo}-spin dynamics. We are interested in particular
in how non-unitary effects in \emph{pure}-spin dynamics are translated into
the corresponding unitary ones in \emph{pseudo}-spin dynamics and vice
versa. We show that as long as there is no magnetic field inhomogeneity the 
\emph{pure}-spin dynamics is unitary, but in the presence of magnetic field
inhomogeneity this dynamics is non-unitary

The paper is organized as follows. We begin, in Section \ref{sec:pseudo} by
highlighting the differences and relationship between \emph{pseudo}\textit{-}
and \emph{pure}-spin models. Section \ref{sec:example} provides a concrete
illustration in terms of a system of two QDs trapping one electron each. In
it, we examine the role of different interactions in both \emph{pseudo}%
\textit{-} and \emph{pure}-spin dynamics. We first derive the coordinate
part of the Hamiltonian (subsection \ref{coord})\ and the form of the
dipolar interaction (subsection \ref{dipole}). In subsections \ref{B-pure}
and \ref{SO-pure}, respectively, we then present calculations illustrating
effects due to both external magnetic field inhomogeneity and the spin-orbit
interaction in the \emph{pure}-spin model. In subsections \ref{B-pseudo} and %
\ref{universal}, we discuss universal quantum gates in both \emph{pseudo}%
\textit{-} and \emph{pure}-spin models. Subsection \ref{SO-pseudo} presents
our estimates for spin-orbit interaction effects in the \emph{pseudo}\textit{%
-}spin model, and compares these estimates to the results of Ref.~\onlinecite%
{Kav01}. We conclude in Section \ref{sec:conc}.

Atomic units, $\hbar =e=m_{e}=1$, \ $1/c\simeq 1/137$, are used throughout
the paper unless stated otherwise.

\section{\emph{Pseudo-} vs \emph{pure}-spin approaches}

\label{sec:pseudo}

In this section we discuss the relation between the present approach based
on the spin-density matrix and the \emph{pseudo}-spin effective Hamiltonian
approach. The latter is usually developed as a low-energy mapping within the
Hubbard model Hamiltonian of interacting electrons \cite%
{Scarola,Hu,Aue,Hub1,Hub2,Hub3,Tak,Mac}. We do not follow the Hubbard model
since it is highly simplified and neglects many interactions which we would
like to keep here. For Hubbard model analyses in the quantum computation
context, see, e.g., Ref.~\onlinecite{Scarola}.

\subsection{\emph{Pseudo}-spin effective Hamiltonian}

In order to keep the present treatment as simple as possible we restrict
ourselves to the two orbitals approximation used in part I; inclusion of
excited-state orbitals is straightforward. Consider the four
single-occupancy basis states $\left\{ \Phi _{s1},\Phi _{ti},i=1,2,3\right\} 
$, where $\Phi _{s1}$ is a singlet wavefunction with two electrons localized
on different QDs, $A$ and $B$, while $\Phi _{ti}$ are the corresponding
triplet wavefunctions. The two double-occupancy states$\mathrm{\,}\left\{
\Phi _{s2},\Phi _{s3}\right\} $ describe two electrons in a singlet state,
localized on the same QD, $A$ or $B$. The total wavefunction $\Psi _{\mathrm{%
tot}}(t)$ in this basis set takes the form%
\begin{equation*}
\Psi _{\mathrm{tot}}(t)=\sum_{i=1}^{3}\left( a_{si}(t)\Phi
_{si}+a_{ti}(t)\Phi _{ti}\right) ,
\end{equation*}%
where the complex amplitudes $\{a_{si}(t),a_{ti}(t)\}$ define, respectively,
the singlet and triplet states population. In the total Hilbert space, the
state is defined by 11 real parameters [12 real parameters defining $%
\{a_{si}(t),a_{ti}(t)\}$ minus a normalization condition]. The unitary
evolution in the total Hilbert space is described by%
\begin{equation*}
\left( 
\begin{array}{c}
\left\vert a_{s}(t)\right\rangle \\ 
\left\vert a_{t}(t)\right\rangle%
\end{array}%
\right) =\exp \left( -iHt\right) \left( 
\begin{array}{c}
\left\vert a_{s}(0)\right\rangle \\ 
\left\vert a_{t}(0)\right\rangle%
\end{array}%
\right) ,
\end{equation*}%
where $H$ is the total two-electron system Hamiltonian.

Since these basis states are orthonormal, projection operators into the
corresponding subspaces can be written as 
\begin{eqnarray}
P &=&\left\vert \Phi _{s1}\right\rangle \left\langle \Phi _{s1}\right\vert
+\sum_{i=1}^{3}\left\vert \Phi _{ti}\right\rangle \left\langle \Phi
_{ti}\right\vert ,\quad  \notag \\
Q &=&\left\vert \Phi _{s2}\right\rangle \left\langle \Phi _{s2}\right\vert
+\left\vert \Phi _{s3}\right\rangle \left\langle \Phi _{s3}\right\vert ,
\label{5.5}
\end{eqnarray}%
where $Q$ projects onto the double occupancy states. Then, using the method
of projection operators, one obtains the Schr\"{o}dinger (eigenvalue)
equation projected into the $P$-subspace%
\begin{equation}
\left( \mathcal{H}_{\mathrm{eff}}(E)-E\right) P\Psi =0,  \label{5.1}
\end{equation}%
where%
\begin{equation}
\mathcal{H}_{\mathrm{eff}}(E)=PHP+PHQ\frac{1}{E-QHQ}QHP.  \label{5.2}
\end{equation}%
Observe that Eq. (\ref{5.1}) is \emph{exact} but non-linear, and 
has 6 solutions.

Due to interelectron repulsion the double occupancy states are usually
much more energetic than the singly-occupied ones if the electrons are well
localized in QDs. We consider the low-energy physics described by Eq. (\ref%
{5.1}) where the total energy $E$ is near the energies of singly-occupied
states. In general $\mathcal{H}_{\mathrm{eff}}$ is not a Hamiltonian since
it is a function of the energy $E$. However, if the energy gap between the $%
P $- and $Q$-states is large enough, one can expand and approximate%
\begin{eqnarray}
\mathcal{H}_{\mathrm{eff}}(E) &=&\mathcal{H}_{\mathrm{eff}}(\bar{E}%
)+\sum_{n=1}^{\infty }PHQ\frac{\left( \bar{E}-E\right) ^{n}}{\left( \bar{E}%
-QHQ\right) ^{n+1}}QHP  \notag \\
&=&\mathcal{H}_{\mathrm{eff}}(\bar{E})+\mathcal{H}_{\mathrm{eff}}^{(1)}(\bar{%
E})\left( \bar{E}-E\right) +O[\left( \bar{E}-E\right) ^{2}]  \notag \\
&\approx &\mathcal{H}_{\mathrm{eff}}(\bar{E})+\mathcal{H}_{\mathrm{eff}%
}^{(1)}(\bar{E})\left( \bar{E}-E\right) ,  \label{5.4}
\end{eqnarray}%
where $\bar{E}$ is an average energy in the $P$-subspace and $\mathcal{H}_{%
\mathrm{eff}}^{(1)}(\bar{E})=PHQ\left( \bar{E}-QHQ\right) ^{-2}QHP$. Keeping
terms up to the first order in Eq. (\ref{5.4}), the non-linear Eq. (\ref{5.1}%
) can be reduced to a generalized linear equation problem%
\begin{equation}
\left( \mathcal{H}_{\mathrm{eff}}(\bar{E})+\mathcal{H}_{\mathrm{eff}}^{(1)}(%
\bar{E})\bar{E}-(1+\mathcal{H}_{\mathrm{eff}}^{(1)}(\bar{E}))E\right) P\Psi
=0.  \label{5.1+1}
\end{equation}%
Solving Eq. (\ref{5.1+1}), we obtain four low energy solutions; the two
high energy, double-occupancy solutions are lost in this approximation.
Therefore, in the low energy, \emph{pseudo}-spin approximation the state is
described by 7 real parameters.

In the following, we assume $\mathcal{H}_{\mathrm{eff}}^{(1)}(\bar{E})\equiv
0$ for simplicity. The effective Hamiltonian Eq. (\ref{5.4}) can be recast
into a \emph{pseudo}-spin form. Using Eq. (\ref{5.5}) we have%
\begin{eqnarray}
\mathcal{H}_{\mathrm{eff}} &=&\mathcal{H}^{ss}\left\vert \Phi
_{s1}\right\rangle \left\langle \Phi _{s1}\right\vert +\sum_{i,j=1}^{3}%
\mathcal{H}_{ij}^{tt}\left\vert \Phi _{ti}\right\rangle \left\langle \Phi
_{tj}\right\vert  \notag \\
&&+\sum_{i=1}^{3}\left( \mathcal{H}_{i}^{st}\left\vert \Phi
_{s1}\right\rangle \left\langle \Phi _{ti}\right\vert +\mathcal{H}%
_{i}^{ts}\left\vert \Phi _{ti}\right\rangle \left\langle \Phi
_{s1}\right\vert \right) ,  \label{5.6}
\end{eqnarray}%
where%
\begin{eqnarray}
\mathcal{H}^{ss} &=&\left\langle \Phi _{s1}\right\vert \mathcal{H}_{\mathrm{%
eff}}(\bar{E})\left\vert \Phi _{s1}\right\rangle ,\quad \mathcal{H}%
_{ij}^{tt}=\left\langle \Phi _{ti}\right\vert \mathcal{H}_{\mathrm{eff}}(%
\bar{E})\left\vert \Phi _{tj}\right\rangle ,  \notag \\
\mathcal{H}_{i}^{st} &=&\left\langle \Phi _{s1}\right\vert \mathcal{H}_{%
\mathrm{eff}}(\bar{E})\left\vert \Phi _{ti}\right\rangle ,\quad \mathcal{H}%
_{i}^{ts}=\left( \mathcal{H}_{i}^{st}\right) ^{\ast }.  \label{5.6.0}
\end{eqnarray}%
In the second quantization representation, the $P$-subspace basis vectors
take the form%
\begin{eqnarray}
\left\vert \Phi _{s1}\right\rangle &=&\frac{1}{\sqrt{2}}\left( c_{A\uparrow
}^{\dagger }c_{B\downarrow }^{\dagger }-c_{A\downarrow }^{\dagger
}c_{B\uparrow }^{\dagger }\right) \left\vert 0\right\rangle =  \notag \\
&&\frac{1}{\sqrt{2}}\left( \left\vert \uparrow \right\rangle _{A}\otimes
\left\vert \downarrow \right\rangle _{B}-\left\vert \downarrow \right\rangle
_{A}\otimes \left\vert \uparrow \right\rangle _{B}\right) ,  \notag \\
\left\vert \Phi _{t1}\right\rangle &=&c_{A\uparrow }^{\dagger }c_{B\uparrow
}^{\dagger }\left\vert 0\right\rangle =\left\vert \uparrow \right\rangle
_{A}\otimes \left\vert \uparrow \right\rangle _{B},  \notag \\
\left\vert \Phi _{t2}\right\rangle &=&\frac{1}{\sqrt{2}}\left( c_{A\uparrow
}^{\dagger }c_{B\downarrow }^{\dagger }+c_{A\downarrow }^{\dagger
}c_{B\uparrow }^{\dagger }\right) \left\vert 0\right\rangle =  \notag \\
&&\frac{1}{\sqrt{2}}\left( \left\vert \uparrow \right\rangle _{A}\otimes
\left\vert \downarrow \right\rangle _{B}+\left\vert \downarrow \right\rangle
_{A}\otimes \left\vert \uparrow \right\rangle _{B}\right) ,  \notag \\
\left\vert \Phi _{t3}\right\rangle &=&c_{A\downarrow }^{\dagger
}c_{B\downarrow }^{\dagger }\left\vert 0\right\rangle =\left\vert \downarrow
\right\rangle _{A}\otimes \left\vert \downarrow \right\rangle _{B},
\label{5.7}
\end{eqnarray}%
where we introduced \emph{pseudo}-spin states $\left\vert s\right\rangle
_{\alpha }$, $s=\uparrow ,\downarrow $, $\alpha =A,B$, localized near the $A$
and $B$ sites [the term \emph{pseudo}\textit{\ }emphasizes the fact that
these are not really spin states since they depend on the electron orbital
degrees of freedom]. Eqs. (\ref{5.7}) establish a one-to-one correspondence
between 4 basis states $\left\{ \Phi _{s1},\Phi _{ti},i=1,2,3\right\} $ and
4 tensor-product \emph{pseudo}-spin states $\left\vert s\right\rangle
_{\alpha }\otimes \left\vert s^{\prime }\right\rangle _{\beta }$, where $%
s,s^{\prime }=\uparrow ,\downarrow $, $\alpha ,\beta =A,B$. Then, relabeling
the \emph{pseudo}-spin states as $\left\vert 0,1\right\rangle =\left\vert
\uparrow ,\downarrow \right\rangle $ and introducing the \emph{pseudo}-spin
Pauli and identity operators%
\begin{eqnarray}
\sigma _{x} &=&\left\vert 0\right\rangle \left\langle 1\right\vert
+\left\vert 1\right\rangle \left\langle 0\right\vert ,  \notag \\
\sigma _{y} &=&-i\left( \left\vert 0\right\rangle \left\langle 1\right\vert
-\left\vert 1\right\rangle \left\langle 0\right\vert \right) ,  \notag \\
\sigma _{z} &=&\left\vert 0\right\rangle \left\langle 0\right\vert
-\left\vert 1\right\rangle \left\langle 1\right\vert ,  \notag \\
I &=&\left\vert 0\right\rangle \left\langle 0\right\vert +\left\vert
1\right\rangle \left\langle 1\right\vert  \label{5.8}
\end{eqnarray}%
where we temporarily dropped the subscripts $A$ and $B$, one easily finds
that%
\begin{eqnarray}
\left\vert \Phi _{s1}\right\rangle \left\langle \Phi _{s1}\right\vert &=&%
\mathbf{S},\quad \left\vert \Phi _{ti}\right\rangle \left\langle \Phi
_{tj}\right\vert =\mathbf{T}_{ij},  \notag \\
\left\vert \Phi _{s1}\right\rangle \left\langle \Phi _{ti}\right\vert &=&%
\mathbf{K}_{i}.  \label{5.9}
\end{eqnarray}%
Here the \emph{pseudo}-spin operators $\mathbf{S}$, $\mathbf{T}_{ij}$ are
defined as 
\begin{equation}
\begin{array}{lll}
\mathbf{S} & = & \frac{1}{4}I-\vec{s}_{A}\cdot \vec{s}_{B},\quad \mathbf{T}%
_{11}=\frac{1}{4}I+\frac{1}{2}S_{z}+s_{Az}s_{Bz}, \\ 
\mathbf{T}_{22} & = & \frac{1}{4}I+s_{Ax}s_{Bx}+s_{Ay}s_{By}-s_{Az}s_{Bz},
\\ 
\mathbf{T}_{33} & = & \frac{1}{4}I-\frac{1}{2}S_{z}+s_{Az}s_{Bz}, \\ 
\mathbf{T}_{12} & = & \frac{1}{\sqrt{2}}\left[ \frac{1}{2}S_{+}+J_{s}\right]
,\quad \mathbf{T}_{23}=\frac{1}{\sqrt{2}}\left[ \frac{1}{2}S_{+}-J_{s}\right]
, \\ 
\mathbf{T}_{13} & = & s_{Ax}s_{Bx}-s_{Ay}s_{By}+i\left(
s_{Ax}s_{By}+s_{Bx}s_{Ay}\right) , \\ 
\mathbf{T} & \mathbf{=} & \frac{3}{4}I+\vec{s}_{A}\cdot \vec{s}_{B},\quad 
\mathbf{T}_{21}=\mathbf{T}_{12}^{\dagger }, \\ 
\mathbf{T}_{31} & = & \mathbf{T}_{13}^{\dagger },\quad \mathbf{T}_{32}=%
\mathbf{T}_{23}^{\dagger }%
\end{array}
\label{5.9+1}
\end{equation}%
where%
\begin{equation*}
\begin{array}{lll}
J_{s} & = & s_{Az}s_{Bx}+s_{Ax}s_{Bz}+i\left(
s_{Az}s_{By}+s_{Ay}s_{Bz}\right) , \\ 
S_{\pm } & = & S_{x}\pm iS_{y},\quad \vec{S}=\vec{s}_{A}+\vec{s}_{B},%
\end{array}%
\end{equation*}%
and $\mathbf{K}$ is defined as
\begin{eqnarray}
\mathbf{K}_{1} &=&-\frac{i}{2\sqrt{2}}\left\{ \left( \vec{J}_{as}\right)
_{x}-i\left( \vec{J}_{as}\right) _{y}\right\} ,\quad \mathbf{K}_{2}=\frac{i}{%
2}\left( \vec{J}_{as}\right) _{z},  \notag \\
\mathbf{K}_{3} &=&\frac{i}{2\sqrt{2}}\left\{ \left( \vec{J}_{as}\right)
_{x}+i\left( \vec{J}_{as}\right) _{y}\right\}  \label{5.9+2}
\end{eqnarray}%
where $\vec{J}_{as}=\left[ \vec{s}_{B}-\vec{s}_{A}\times \vec{S}\right] $.
In fact, Eqs. (\ref{5.9+1}) and (\ref{5.9+2}) can be obtained from the
corresponding ones in part I if the \emph{pure}-spin operators $\vec{s}%
_{1,2} $ are replaced respectively by the \emph{pseudo}-ones, $\vec{s}_{A,B}$%
. We reproduce these formulas here in order to make the presentation as
self-contained as possible.

As is seen from Eqs. (\ref{5.9+1}) and (\ref{5.9+2}), the first line of Eq. (%
\ref{5.6}) is symmetric with respect to spin permutations [$A\leftrightarrow
B$], while the second one is asymmetric representing, in particular, the
Dzyaloshinskii-Moriya (DM-type) interaction term.\cite{Dzy,Mor} Notice that
these asymmetric (in spin permutations) terms cancel out of unitary spin
dynamics after averaging over orbital degrees of freedom, as demonstrated in
part I. However, they do not disappear completely, but rather are converted
into the corresponding non-unitary terms plus the \emph{Lamb shift} term as
will be seen in next subsection. From the symmetric part of the Hamiltonian~(\ref{5.6}), using Eqs. (\ref{5.9+1}) one can derive the isotropic
Heisenberg exchange interaction term%
\begin{equation}
\mathcal{H}_{H}=J_{H}\vec{s}_{A}\cdot \vec{s}_{B},  \label{5.10}
\end{equation}%
where the Heisenberg exchange interaction constant $J_{H}=\frac{1}{3}\sum_{i}%
\mathcal{H}_{ii}^{tt}-\mathcal{H}^{ss}$; in contrast, as was demonstrated in
part I, the Heisenberg interaction term does not affect the unitary
evolution of the spin density matrix, apart from the Lamb-energy
shift. In subsection \ref{B-pure}, we demonstrate numerically the effects of
the Heisenberg interaction on both the Lamb-energy shift and the
non-unitary part of the spin density matrix evolution.

Observe that the asymmetric part of the Hamiltonian Eq. (\ref{5.6}) is
proportional to the singlet-triplet subspace interaction matrix $\mathcal{H}%
_{i}^{st}$, which is responsible for the coupling between singlet and
triplet states. As will be demonstrated in Section \ref{sec:example}, the
non-zero coupling between these states is due to $\vec{B}$-field spatial
inhomogeneity (i.e., it cannot arise due to the homogeneous component of the
external magnetic field), as well as due to the spin-orbit interaction.

\subsection{Spin density matrix}

In part I we derived the Lindblad-type master equation for the spin density
matrix

\begin{eqnarray}
\frac{\partial \rho (t)}{\partial t} &=&-i\left[ \mathbf{\tilde{H}}_{\alpha
}^{tt},\rho (t)\right] +\mathcal{L}_{\alpha }[\rho (t)],  \label{Leq1} \\
\mathbf{\tilde{H}}_{\alpha }^{tt} &=&\sum_{ij}\left( H^{tt}+\frac{1}{2}%
P_{\alpha }\right) _{ij}\mathbf{T}_{ij}=\mathbf{H}^{tt}+\frac{1}{2}\mathbf{P}%
_{\alpha },  \notag \\
\mathcal{L}_{\alpha }[\rho (t)] &=&\frac{1}{2}\sum_{ij}(\chi _{\alpha
})_{ij}\left( \left[ \mathbf{K}_{i},\rho (t)\mathbf{K}_{j}^{\dagger }\right]
+\left[ \mathbf{K}_{i}\rho (t),\mathbf{K}_{j}^{\dagger }\right] \right) , 
\notag
\end{eqnarray}%
where the first and second terms describe, respectively, unitary and
non-unitary contributions to the evolution. $\mathbf{\tilde{H}}_{\alpha
}^{tt}$ is an effective \emph{pure}-spin Hamiltonian which includes the 
Lamb-shift term, $\frac{1}{2}\mathbf{P}_{\alpha }$; the \emph{pure}%
-spin operators $\mathbf{T}_{ij}$ and $\mathbf{K}_{i}$ are defined by Eqs. (%
\ref{5.9+1}) and (\ref{5.9+2}) where $\vec{s}_{A,B}\rightarrow \vec{s}_{1,2}$%
. The index $\alpha =\{s,t,m\}$ specifies which initial state $\rho (0)$,
singlet, $s$, triplet, $t$, or a mixed one, $m$, is taken.

As mentioned in part I, all the matrix functions in Eq. (\ref{Leq1}) as well
as the \emph{pseudo}-spin Hamiltonian Eq. (\ref{5.6}) are expressible in
terms of $H$ matrix elements%
\begin{eqnarray}
H &=&\left( 
\begin{array}{cc}
H^{ss} & H^{st} \\ 
H^{ts} & H^{tt}%
\end{array}%
\right) ,  \label{Leq2} \\
H_{ij}^{\alpha \beta } &=&\left\langle \Phi _{\alpha i}\right\vert
H\left\vert \Phi _{\beta j}\right\rangle ,\quad \alpha ,\beta =s,t,\quad
i,j=1,2,3.  \notag
\end{eqnarray}

In the following example, we consider the triplet case for which we have

\begin{eqnarray}
\chi _{\alpha }^{T} &=&i\left( Q_{\alpha }-Q_{\alpha }^{\dagger }\right) ,
\label{Leq3} \\
P_{\alpha } &=&Q_{\alpha }+Q_{\alpha }^{\dagger }  \notag
\end{eqnarray}%
where%
\begin{equation}
\begin{array}{ll}
Q_{\alpha }= & \sum\limits_{k}\exp (-i\varepsilon _{k}t)H^{ts}\left[
\left\vert e_{sk}\right\rangle \left\langle e_{sk}\right\vert R_{\alpha
}(0)+\left\vert e_{sk}\right\rangle \left\langle e_{tk}\right\vert \right]
\times \\ 
& \left( \sum\limits_{k}\exp (-i\varepsilon _{k}t)\left[ \left\vert
e_{tk}\right\rangle \left\langle e_{sk}\right\vert R_{\alpha }(0)+\left\vert
e_{tk}\right\rangle \left\langle e_{tk}\right\vert \right] \right) ^{-1}.%
\end{array}
\label{Leq4}
\end{equation}%
Here, $R_{\alpha }(0)$ \ is a correlation matrix, which establishes an
initial correlation between the singlet and triplet amplitudes 
\begin{equation}
a_{s}(0)=R_{\alpha }(0)a_{t}(0)  \label{Leq4+1}
\end{equation}%
[in the triplet case, we have $R_{\alpha =t}(0)\equiv 0$; in the mixed case,
where both $a_{s}(0)\neq 0$ and $a_{t}(0)\neq 0$, $R_{\alpha =m}(0)\neq 0$;
for the singlet case, see part I] and$\ \varepsilon _{k}$, $\left\vert
e_{sk}\right\rangle $, and $\left\vert e_{tk}\right\rangle $ are solutions
to the eigenvalue problem

\begin{equation}
\left( 
\begin{array}{cc}
H^{ss} & H^{st} \\ 
H^{ts} & H^{tt}%
\end{array}%
\right) \left( 
\begin{array}{c}
\left\vert e_{sk}\right\rangle \\ 
\left\vert e_{tk}\right\rangle%
\end{array}%
\right) =\varepsilon _{k}\left( 
\begin{array}{c}
\left\vert e_{sk}\right\rangle \\ 
\left\vert e_{tk}\right\rangle%
\end{array}%
\right) ,\quad k=1,\cdots ,6.  \label{Leq5}
\end{equation}

\section{Example:\ System of two quantum dots}

\label{sec:example}

In this section, we investigate the role of different interactions in the
calculation of the $H$ matrix. Let us consider a system of two electrons
trapped at sites $\vec{r}_{A}$ and $\vec{r}_{B}$ ($\vec{r}_{A,B}$ are
radius-vectors of the centers of QDs in the $z=0$ plane) created by a system
of charged electrodes in a semiconductor \ heterostructure so that the
electrons are confined in the $z=0$ plane or a system of localized
conduction-band electrons in $n$-doped GaAs as in our calculation example.
The heterostructure trapping potential

\begin{equation}
V_{\mathrm{tr}}(z,\vec{r})=V_{\perp }(z)+V_{A}(\vec{r})+V_{B}(\vec{r})
\label{Eq60}
\end{equation}%
is separable in the in-plane and out-of-plane directions; $V_{\perp }(z)$
and $V_{A,B}(\vec{r})$ are the trapping potentials in the $z$-direction and
in the $z=0$ plane around $\vec{r}_{A,B}$ respectively. If the electron
system is placed in a constant magnetic field $\vec{B}_{0}$ directed along
the $z$-axis (with vector potential $\vec{A}_{0}=\frac{1}{2}\left[ \vec{r}%
\times \vec{B}_{0}\right] $), then the in-plane motion, in a superposition
of the\ in-plane confining oscillatory potential and a perpendicular
magnetic field, is described by the Fock-Darwin (FD) states.\cite{Jacak}
Approximating the confining potential by a quadratic one 
\begin{equation}
V_{A,B}(\vec{r})\approx \frac{1}{2}\omega _{A,B}^{2}\left( \vec{r}-\vec{r}%
_{A,B}\right) ^{2},  \label{Eq61}
\end{equation}%
we can take as basis \textquotedblleft atomic\textquotedblright\ orbitals
the ground-state functions 
\begin{equation}
\phi _{A,B}(z,\vec{r})=\varphi _{0}(z)R_{A,B}^{\mathrm{FD}}(|\vec{r}-\vec{r}%
_{A,B}|),  \label{Eq69}
\end{equation}%
where the out-of-plane motion in the $z$-direction is \textquotedblleft
frozen\textquotedblright\ in the ground state $\varphi _{0}(z)$ in the
potential $V_{\perp }(z)$, and the ground FD state is%
\begin{eqnarray}
R_{A,B}^{\mathrm{FD}} &=&\frac{1}{\sqrt{2\pi }l_{A,B}}\exp \left( -\frac{%
r^{2}}{4l_{A,B}^{2}}\right) ,\quad  \label{Eq64} \\
l_{A,B} &=&\frac{l_{c}}{\sqrt[4]{1+4\omega _{A,B}^{2}/\omega _{c}^{2}}}%
,\quad l_{c}=\sqrt{\frac{c}{B_{0}}}.  \notag
\end{eqnarray}%
Here $l_{A,B}$ is the effective length scale, equal to the magnetic length $%
l_{c}$ in the absence of the confining potential, $\omega _{A,B}\equiv 0$; $%
\omega _{c}=B_{0}/c$ is the cyclotron frequency.

The orbitals Eq. (\ref{Eq69}) must be orthogonalized. One way to do this is\
a simple Gram-Schmidt orthogonalization procedure:%
\begin{equation}
\begin{array}{ll}
\tilde{\phi}_{A}= & \phi _{A}, \\ 
\tilde{\phi}_{B}= & \frac{1}{\sqrt{1-S^{2}}}\left( \phi _{B}-S\phi
_{A}\right) ,%
\end{array}
\label{Eq70}
\end{equation}%
where the overlap matrix element $S_{AB}=S_{BA}=S=\left\langle \phi
_{A}\right. \left\vert \phi _{B}\right\rangle $ can be calculated
analytically%
\begin{equation}
S=\frac{2l_{A}l_{B}}{l_{A}^{2}+l_{B}^{2}}\exp \left( -\frac{r_{AB}^{2}}{%
4(l_{A}^{2}+l_{B}^{2})}\right) .  \label{Eq71}
\end{equation}%
For appropriate values of system parameters such as the interdot distance $%
r_{AB}$ and the external magnetic field $B_{0}$, the overlap becomes
exponentially small.

The other, more symmetric way is to make a transition to the
\textquotedblleft molecular\textquotedblright\ or two-centered orbitals by
pre-diagonalizing the coordinate part of Pauli's \ non-relativistic
Hamiltonian $\hat{h}_{c}$, which describes the electron's motion in a
superposition of the trapping potential and magnetic fields:%
\begin{equation}
\begin{array}{ccc}
\tilde{\phi}_{A} & = & c_{AA}\phi _{A}+c_{AB}\phi _{B}, \\ 
\tilde{\phi}_{B} & = & c_{BA}\phi _{A}+c_{BB}\phi _{B}, \\ 
\left\langle \tilde{\phi}_{i}\right\vert \left. \tilde{\phi}_{j}\right\rangle
& = & \delta _{ij},\qquad i,j=A,B, \\ 
\left\langle \tilde{\phi}_{i}\right\vert \hat{h}_{c}\left\vert \tilde{\phi}%
_{j}\right\rangle & = & \varepsilon _{i}\delta _{ij},\qquad i,j=A,B.%
\end{array}
\label{Eq71.1}
\end{equation}%
The two-state eigenvalue problem Eq. (\ref{Eq71.1}) is solved analytically
in terms of \textquotedblleft atomic\textquotedblright\ orbitals matrix
elements: $h_{ij}=\left\langle \phi _{i}\right\vert \hat{h}_{c}\left\vert
\phi _{j}\right\rangle $.

In general, given the \textquotedblleft molecular\textquotedblright\ Eq. (%
\ref{Eq71.1}) or \textquotedblleft half-molecular\textquotedblright\ Eq. (%
\ref{Eq70}) basis choices, one cannot ascribe a spin to a particular
QD, since an electron in a molecular orbital belongs to both QDs.

The total Hamiltonian contains both coordinate and spin-dependent terms.
First we consider the coordinate part of the Hamiltonian in the $\tilde{\phi}%
_{A,B}$ basis set.

\subsection{Coordinate part of the Hamiltonian}

\label{coord}

In view of the orthogonality of the singlet and triplet spin wave functions,
the spin-independent part of the Hamiltonian does not contribute to the
singlet-triplet coupling, $H_{c}^{st}=H_{c}^{ts}=0$, whereas for the
singlet-singlet and triplet-triplet Hamiltonians we get%
\begin{eqnarray}
H_{c}^{ss} &=&\{H_{cij}^{ss\,}\}_{i,j=1}^{3},\quad
H_{cij}^{ss}=H_{cji}^{ss\ast }  \notag \\
H_{c11}^{ss} &=&\tilde{h}_{AA}+\tilde{h}_{BB}+\tilde{v}_{\mathrm{ee}}(AB;AB)+%
\tilde{v}_{\mathrm{ee}}(AB;BA),  \notag \\
H_{c12}^{ss} &=&\sqrt{2}\left( \tilde{h}_{BA}+\tilde{v}_{\mathrm{ee}%
}(AB;AA)\right) ,\quad  \label{Eq72} \\
H_{c13}^{ss} &=&\sqrt{2}\left( \tilde{h}_{AB}+\tilde{v}_{\mathrm{ee}%
}(AB;BB)\right) ,\quad  \notag \\
H_{c22}^{ss} &=&2\tilde{h}_{AA}+\tilde{v}_{\mathrm{ee}}(AA;AA),\quad
H_{c23}^{ss}=\tilde{v}_{\mathrm{ee}}(AA;BB),  \notag \\
H_{c33}^{ss} &=&2\tilde{h}_{BB}+\tilde{v}_{\mathrm{ee}}(BB;BB),  \notag
\end{eqnarray}%
\begin{eqnarray}
H_{c}^{tt} &=&\varepsilon _{t}I,\quad  \label{Eq73} \\
\varepsilon _{t} &=&\tilde{h}_{AA}+\tilde{h}_{BB}+\tilde{v}_{\mathrm{ee}%
}(AB;AB)-\tilde{v}_{\mathrm{ee}}(AB;BA),  \notag
\end{eqnarray}%
where 
\begin{eqnarray}
\tilde{h}_{ij} &=&\left\langle \tilde{\phi}_{i}\right\vert \hat{h}%
_{c}\left\vert \tilde{\phi}_{j}\right\rangle ,\quad i,j=A,B  \notag \\
\tilde{v}_{\mathrm{ee}}(ij;kl) &=&\left\langle \tilde{\phi}_{i}(1)\tilde{\phi%
}_{j}(2)\right\vert \frac{1}{\varepsilon r_{12}}\left\vert \tilde{\phi}%
_{k}(1)\tilde{\phi}_{l}(2)\right\rangle ,\quad  \label{Eq74} \\
i,j,k,l &=&A,B  \notag
\end{eqnarray}%
with $\tilde{h}_{ij}=\varepsilon _{i}\delta _{ij}$ for \textquotedblleft
molecular\textquotedblright\ orbitals and $\tilde{v}_{\mathrm{ee}}$ being
the interelectron electrostatic interaction matrix elements. The matrix $%
H_{c}^{ss}$ is diagonally dominated if the overlap $S\ll 1$; $H_{c11}^{ss}$
is the singlet energy of the singly occupied state whereas $H_{c22}^{ss}$
and $H_{c33}^{ss}$ are energies of doubly occupied states if one neglects
the coupling between single- and double-occupancy states. Observe that the
Heisenberg constant $J_{H}=\varepsilon _{t}-\varepsilon _{s}$ where $%
\varepsilon _{s}$ is the lowest eigenvalue of the matrix $H_{c}^{ss}$.\ The
matrix elements $\tilde{h}_{ij}$ and $\tilde{v}_{\mathrm{ee}}(ij;kl)$ can
trivially be expressed in terms of the corresponding matrix elements $h_{ij}$
and $v_{\mathrm{ee}}(ij;kl)$ where the orthonormalized states $\tilde{\phi}%
_{i}$'s are replaced by $\phi _{i}$'s using the relations Eq. (\ref{Eq70})
or (\ref{Eq71.1}). \ 

\subsection{Dipole spin-spin interaction}

\label{dipole}

In the total spin representation, the dipole spin-spin interaction can be
rewritten as \cite{Messiah}%
\begin{eqnarray}
V_{\mathrm{dip}} &=&\frac{1.45}{2}\,\mathrm{meV}\left( \frac{%
S^{2}\,r_{12}^{2}-3(\vec{S}\cdot \vec{r}_{12})^{2}}{r_{12}^{5}}-\right.  
\notag \\
&&\left. \frac{8\pi }{3}\left( S^{2}-\frac{3}{2}\right) \,\delta (\vec{r}%
_{12})\right) .  \label{Eq78}
\end{eqnarray}%
Since $\vec{S}\left\vert \chi _{s}\right\rangle =0$ and $f_{t}(\vec{r}_{1}=%
\vec{r}_{2})=0$, where $\left\vert \chi _{s}\right\rangle $ and $f_{t}$ are
singlet-state spin and triplet-state coordinate wavefunctions, we have $H_{%
\mathrm{dip}}^{st}=H_{\mathrm{dip}}^{ts}=0$ and a non-zero contribution to $%
H_{\mathrm{dip}}^{ss}$ comes only from the contact term:%
\begin{eqnarray}
(H_{\mathrm{dip}}^{ss})_{ij} &=&(1.45\cdot 2\pi )\mathrm{\ meV}\left\langle
f_{si}\right\vert \delta (\vec{r}_{12})\left\vert f_{sj}\right\rangle  
\notag \\
&=&1.09\Lambda \mathrm{\ meV}\left( 
\begin{array}{ccc}
d_{11} & d_{12} & d_{13} \\ 
d_{12} & d_{22} & \frac{1}{2}d_{11} \\ 
d_{13} & \frac{1}{2}d_{11} & d_{33}%
\end{array}%
\right)   \label{Eq79}
\end{eqnarray}%
where $\Lambda $ is an effective constant of the interaction that confines
electrons in the $z$-plane and 
\begin{eqnarray}
d_{11} &=&2\left\langle \tilde{\phi}_{A}^{2}\right. \left\vert \tilde{\phi}%
_{B}^{2}\right\rangle =\frac{1}{l_{A}^{2}+l_{B}^{2}}\exp \left( -\frac{%
r_{AB}^{2}}{2(l_{A}^{2}+l_{B}^{2})}\right) ,  \notag \\
d_{12} &=&\sqrt{2}\left\langle \tilde{\phi}_{A}^{3}\right. \left\vert \tilde{%
\phi}_{B}\right\rangle =\sqrt{2}\frac{l_{B}/l_{A}}{3l_{B}^{2}+l_{A}^{2}}\exp
\left( -\frac{3}{4}\frac{r_{AB}^{2}}{3l_{B}^{2}+l_{A}^{2}}\right) ,  \notag
\\
d_{13} &=&\sqrt{2}\left\langle \tilde{\phi}_{A}\right. \left\vert \tilde{\phi%
}_{B}^{3}\right\rangle =\sqrt{2}\frac{l_{A}/l_{B}}{3l_{A}^{2}+l_{B}^{2}}\exp
\left( -\frac{3}{4}\frac{r_{AB}^{2}}{3l_{A}^{2}+l_{B}^{2}}\right) ,  \notag
\\
d_{22} &=&\left\langle \tilde{\phi}_{A}^{2}\right. \left\vert \tilde{\phi}%
_{A}^{2}\right\rangle =\frac{1}{4l_{A}^{2}},\quad d_{33}=\left\langle \tilde{%
\phi}_{B}^{2}\right. \left\vert \tilde{\phi}_{B}^{2}\right\rangle =\frac{1}{%
4l_{B}^{2}}.  \label{Eq80}
\end{eqnarray}%
The magnetic dipole contribution to the triplet-triplet interaction
Hamiltonian can be written as%
\begin{equation}
H_{\mathrm{dip}}^{tt}=0.36\,\mathrm{meV}\left( 
\begin{array}{ccc}
\bar{t}_{0} & -\frac{3}{\sqrt{2}}\bar{t}_{1}^{\ast } & -3\bar{t}_{2}^{\ast }
\\ 
-\frac{3}{\sqrt{2}}\bar{t}_{1} & -2\bar{t}_{0} & \frac{3}{\sqrt{2}}\bar{t}%
_{1}^{\ast } \\ 
-3\bar{t}_{2} & \frac{3}{\sqrt{2}}\bar{t}_{1} & \bar{t}_{0}%
\end{array}%
\right)   \label{Eq81}
\end{equation}%
where $t_{i},$ $i=0,1,2$ are dipole tensor operators%
\begin{eqnarray}
t_{0} &=&\frac{1-3\cos ^{2}\theta _{12}}{r_{12}^{3}},\quad   \notag \\
t_{1} &=&\frac{\sin 2\theta _{12}\exp (i\varphi _{12})}{r_{12}^{3}},\quad 
\label{Eq82} \\
t_{2} &=&\frac{\sin ^{2}\theta _{12}\exp (2i\varphi _{12})}{r_{12}^{3}} 
\notag
\end{eqnarray}%
with $(r_{12},\theta _{12},\varphi _{12})$ being spherical coordinates of
the interelectron radius-vector $\vec{r}_{12}=$ $\vec{r}_{1}-\vec{r}_{2}$;
the bar over $t_{i}$ denotes averaging over the triplet coordinate
wavefunction:%
\begin{equation}
\bar{t}_{i}=\int \int d^{3}\vec{r}_{1}d^{3}\vec{r}_{2}\left\vert f_{t}(\vec{r%
}_{1},\vec{r}_{2})\right\vert ^{2}t_{i}(\vec{r}_{1},\vec{r}_{2})
\label{Eq83}
\end{equation}

Taking into account the fact that the electrons are exponentially localized
at sites $\vec{r}_{A}$ and $\vec{r}_{B}$ in the $f_{t}$ state, a good
approximation to $\bar{t}_{i}$ is to approximate the function $t_{i}$ by a
constant value at those points where $f_{t}(\vec{r}_{1},\vec{r}_{2})$ is
localized, thus obtaining the estimate%
\begin{equation}
H_{\mathrm{dip}}^{tt}=\frac{0.36}{r_{AB}^{3}}\,\mathrm{meV}\left( 
\begin{array}{ccc}
1 & 0 & -3 \\ 
0 & -2 & 0 \\ 
-3 & 0 & 1%
\end{array}%
\right) .  \label{Eq84}
\end{equation}%
In order to further improve the estimate, the function $t_{i}(\vec{r}_{1},%
\vec{r}_{2})$ can be expanded in a Taylor's series around the localization
points and the remaining integrals in the expansion terms be calculated
analytically. From \ Eq. (\ref{Eq84}) we find the estimate $H_{\mathrm{dip}%
}^{tt}\sim (0.36/r_{AB}^{3})\,\mathrm{meV}\approx 5.0\cdot 10^{-8}\,\mathrm{%
meV}$ at $r_{AB}=100\,\mathring{A}$.

\subsection{The $\vec{B}$-field interaction in the \emph{pure}-spin model}

\label{B-pure}

For the magnetic field one gets%
\begin{equation}
H^{tt}(\vec{B})=\left( 
\begin{array}{ccc}
B_{avz} & B_{av}^{-} & 0 \\ 
B_{av}^{+} & 0 & B_{av}^{-} \\ 
0 & B_{av}^{+} & -B_{avz}%
\end{array}%
\right) ,  \label{Eq85}
\end{equation}%
where%
\begin{eqnarray}
\vec{B}_{av} &=&\frac{1}{2}\left( \left\langle \tilde{\phi}_{A}\right\vert 
\vec{B}\left\vert \tilde{\phi}_{A}\right\rangle +\left\langle \tilde{\phi}%
_{B}\right\vert \vec{B}\left\vert \tilde{\phi}_{B}\right\rangle \right)
,\qquad  \notag \\
B_{av}^{\pm } &=&\frac{1}{\sqrt{2}}\left( B_{avx}\pm iB_{avy}\right) .
\label{Eq85.1}
\end{eqnarray}

Using Eqs. (\ref{5.9+1}) and (\ref{Eq85}), one derives the Zeeman interaction
Hamiltonian of the total spin $\vec{S}$ with the magnetic field $\vec{B}%
_{av} $: 
\begin{equation}
\mathbf{H}^{tt}(\vec{B})=\sum_{ij=1}^{3}H_{ij}^{tt}(\vec{B})\,\mathbf{T}%
_{ij}=\vec{B}_{av}\cdot \vec{S}.  \label{Eq86}
\end{equation}

Similarly, for the singlet-triplet matrix we have%
\begin{equation}
H^{st}(\vec{B})=\left( 
\begin{array}{ccc}
\frac{1}{2\sqrt{2}}\Delta B^{+} & -\frac{1}{2}\Delta B_{z} & -\frac{1}{2%
\sqrt{2}}\Delta B^{-} \\ 
\frac{1}{2}\delta B^{+} & -\frac{1}{\sqrt{2}}\delta B_{z} & -\frac{1}{2}%
\delta B^{-} \\ 
-\frac{1}{2}\delta B^{-\ast } & \frac{1}{\sqrt{2}}\delta B_{z}^{\ast } & 
\frac{1}{2}\delta B^{+\ast }%
\end{array}%
\right) ,  \label{Eq87}
\end{equation}%
where%
\begin{eqnarray}
\Delta \vec{B} &=&\left\langle \tilde{\phi}_{B}\right\vert \vec{B}\left\vert 
\tilde{\phi}_{B}\right\rangle -\left\langle \tilde{\phi}_{A}\right\vert \vec{%
B}\left\vert \tilde{\phi}_{A}\right\rangle ,\qquad  \notag \\
\Delta B^{\pm } &=&\Delta B_{x}\pm i\Delta B_{y},  \notag \\
\delta \vec{B} &=&\left\langle \tilde{\phi}_{A}\right\vert \vec{B}\left\vert 
\tilde{\phi}_{B}\right\rangle ,\qquad \delta B^{\pm }=\delta B_{x}\pm
i\delta B_{y}.  \label{Eq87.1}
\end{eqnarray}%
If the $\vec{B}$-field is homogeneous, from Eq. (\ref{Eq87.1}) we obtain $%
\Delta \vec{B}=\delta \vec{B}=0$ and $H^{st}(\vec{B})=0$. In this case, the
spin dynamics is unitary and is described by the Zeeman Hamiltonian $\mathbf{%
H}^{tt}(\vec{B})$ Eq. (\ref{Eq86}); the spin-spin dipole interaction $H_{%
\mathrm{dip}}^{tt}$ is too small and can usually be safely neglected.

Let us consider\ modifications due the to $\vec{B}$-field inhomogeneity in
the \emph{pure}-spin model. Neglecting contributions from the
double-occupancy states within the first-order perturbation approximation in
the singlet-triplet interaction $H^{st}$, we find for the non-unitary term
in Eq. (\ref{Leq1})%
\begin{eqnarray}
\mathcal{L}_{t} &=&\frac{1}{2}\sum_{ij}(\chi _{t})_{ij}\left( \left[ \mathbf{%
K}_{i},\rho \mathbf{K}_{j}^{\dagger }\right] +\left[ \mathbf{K}_{i}\rho ,%
\mathbf{K}_{j}^{\dagger }\right] \right)  \notag \\
&=&\frac{\sin (J_{H}t)}{J_{H}}\left( \left[ \mathbf{K},\rho \mathbf{K}%
^{\dagger }\right] +\left[ \mathbf{K}\rho ,\mathbf{K}^{\dagger }\right]
\right) ,  \label{Eq88}
\end{eqnarray}%
where%
\begin{equation}
\mathbf{K=}\sum_{i}H_{1i}^{st}(\vec{B})\mathbf{K}_{i}=-\frac{i}{4}\Delta 
\vec{B}\cdot \vec{J}_{\mathrm{as}}  \label{Eq88.1}
\end{equation}%
and%
\begin{equation}
\vec{J}_{\mathrm{as}}=\left[ \vec{s}_{2}-\vec{s}_{1}\times \vec{S}\right] =2%
\left[ \vec{s}_{2}\times \vec{s}_{1}\right] +i(\vec{s}_{2}-\vec{s}_{1})
\label{Eq88.2}
\end{equation}%
is an asymmetric spin operator containing both linear and bilinear parts.
Observe that $\mathcal{L}_{t}=0$ at the \textquotedblleft
swap\textquotedblright\ times $t_{n}=\pi n/J_{H}$, $n=0,1,\ldots $.

Similarly, for the \textit{Lamb} \textit{shift} \ in Eq. (\ref{Leq1}) we have%
\begin{equation}
\mathbf{L}_{t}\mathbf{=}\frac{1}{2}\sum_{ij}(P_{t})_{ij}\mathbf{K}%
_{i}^{\dagger }\mathbf{K}_{j}=\frac{1-\cos (J_{H}t)}{J_{H}}\mathbf{K}%
^{\dagger }\mathbf{K.}  \label{Eq89}
\end{equation}%
Observe that $\mathcal{L}_{t}$ and $\mathbf{L}_{t}$ are quadratic in the
difference field $\Delta \vec{B}$. Besides, notice that the magnetic field
due to spin-orbit coupling does not contribute to the difference field $%
\Delta \vec{B}_{\mathrm{so}}=0$ but contributes to the $\delta \vec{B}_{%
\mathrm{so}}$-field that is present in the coupling between the triplet
states and the double occupancy, singlet states, in Eq. (\ref{Eq87}). If the
external magnetic field $\vec{B}_{\mathrm{ex}}$ is homogeneous, then the
singlet-triplet states coupling comes only from the spin-orbit interaction.
Since the double-occupancy states should be involved in the dynamics in
order to obtain non-zero spin-orbit interaction effects, these effects are
expected to be especially small, proportional to $\delta B_{\mathrm{so}%
}^{2}, $ in the \textit{pure}-spin model. An estimate of these spin-orbit
effects will be given in a numerical example in the next Section.

\begin{figure}[tbp]
\epsfig{
file=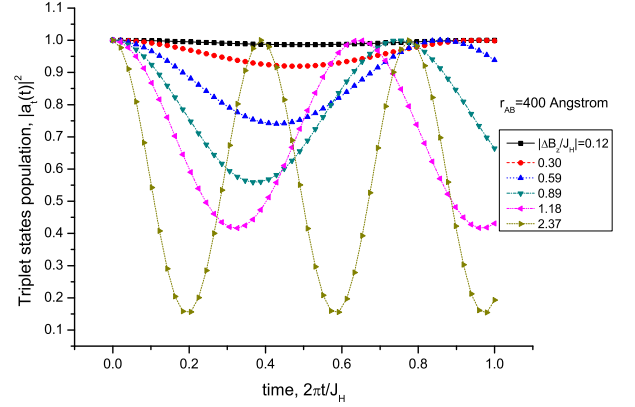,
width=3.5in,height=2.5in} 
\caption{(color online) The triplet
states population of electrons in shallow QD centers in GaAs, as a function
of time at different magnetic field differences $\Delta
B_{z}=0.01,0.025,0.05,0.075,0.1,0.2$T, normalized to the Heisenberg exchange 
$J_{H}$ constant. Initially the system is in the triplet state $\left\vert
S=1,M_{S}=0\right\rangle $. The distance between QDs is 400 \AA .}
\label{Fig1}
\end{figure}

Clearly, there is an important qualitative difference between \emph{pure}%
\textit{-} and \emph{pseudo}\textit{-}spin models. In the former, the
singlet-triplet states coupling is a second order effect, while in the
latter this coupling is of first order in $H^{st}$ [cf., Eq. (\ref{Eq88})
and (\ref{5.6})]. Thus, in \emph{pure}\textit{-}spin models effects due to $%
\vec{B}$-field inhomogeneity should be especially (quadratically) small as
compared to the corresponding \emph{pseudo}\textit{-}spin model effects. In
case of negligible $\vec{B}$-field inhomogeneity, as follows from Eq. (\ref%
{Eq87}),\ the \emph{pure}\textit{-}spin dynamics is unitary and is governed
by the spin Hamiltonian $\mathbf{H}^{tt}$.

\begin{figure}[tbp]
\epsfig{
file=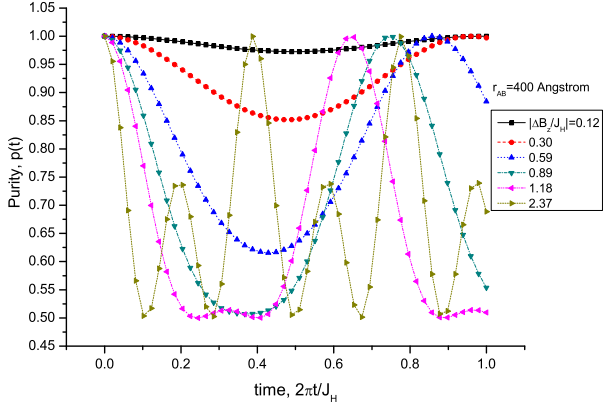,
width=3.5in,height=2.5in} 
\caption{(color online) The purity $\
p(t)=\mathrm{Tr\,}\protect\rho ^{2}(t)$ for the same parameters as in Fig. 1.}
\label{Fig2}
\end{figure}

\begin{figure}[tbp]
\epsfig{
file=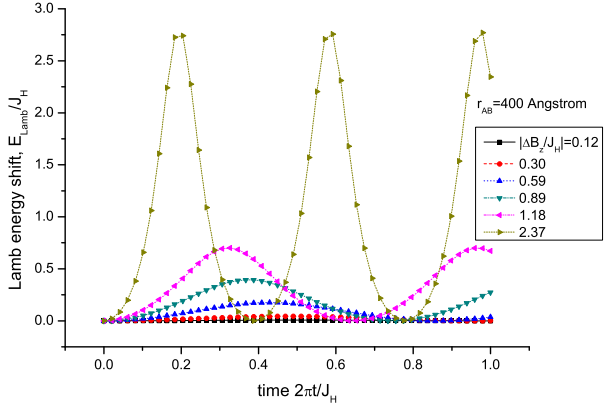,
width=3.5in,height=2.5in} 
\caption{(color online) The Lamb shift
energy as a function of time for the same parameters as in Fig. 1.}
\label{Fig3}
\end{figure}

\begin{figure}[tbp]
\epsfig{
file=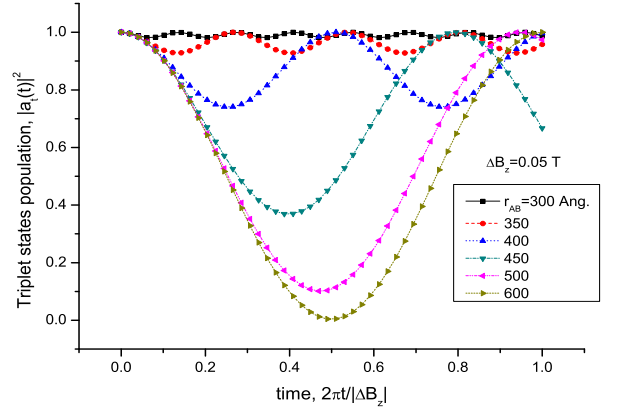,
width=3.5in,height=2.5in} 
\caption{(color online) Triplet states
population dependence on interdot separation $r_{AB}=300,$ $\,350,$ $400,$ $%
\,450,$ $500$, and $600$\AA ,\ at a fixed $\Delta B_{z}=0.05$T.}
\label{Fig4}
\end{figure}

\begin{figure}[tbp]
\epsfig{
file=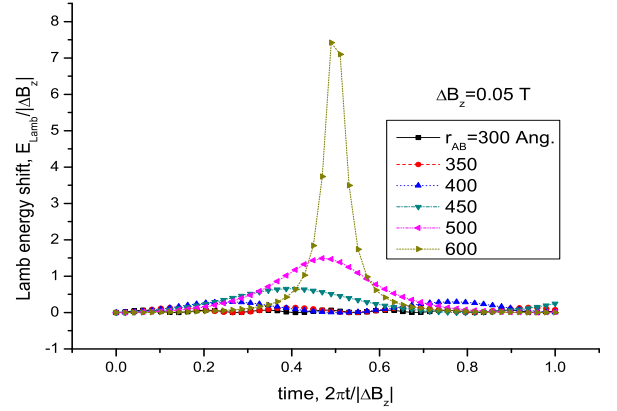,
width=3.5in,height=2.5in} 
\caption{(color online) Lamb energy
shift dependence on interdot distance $r_{AB}=300,$ $\,350,$ $400,$ $\,450,$ 
$500$, and $600$\AA ,\ at a fixed $\Delta B_{z}=0.05$T.}
\label{Fig5}
\end{figure}

Let us now consider a simple numerical example for the non-unitary effects
due to the difference field $\Delta \vec{B}$ for an electron localized on a
donor impurity in an $n$-doped GaAs semiconductor. To simplify numerics, we
assume that $H^{ss}={\rm diag}(\varepsilon _{s1},\varepsilon
_{s2},\varepsilon _{s3})$ is diagonal and the singlet-triplet coupling
field, $\Delta \vec{B}$, has only a non-zero $z$-component, $\Delta B_{z}$.
Then, the corresponding eigenvalue problem, Eq. (\ref{Leq5}), can be reduced
to a biquadratic polynomial equation which could, in principle, be solved
exactly. If we neglect the exponentially small coupling field $\delta B_{z}$%
, proportional to the overlap $S$, the biquadratic equation reduces to a
quadratic one and we find for the non-unitary term%
\begin{eqnarray}
\mathcal{L}_{t} &=&\frac{\omega \sin \omega t}{J_{H}^{2}+\frac{1}{2}\Delta
B_{z}^{2}(1+\cos \omega t)}\times  \notag \\
&&\left( \left[ \mathbf{K},\rho \mathbf{K}^{\dagger }\right] +\left[ \mathbf{%
K}\rho ,\mathbf{K}^{\dagger }\right] \right) \quad  \label{Eq88+1} \\
\mathbf{K} &=&-\frac{i}{4}\Delta B_{z}\cdot J_{asz}  \notag
\end{eqnarray}%
and the \textit{Lamb} \textit{shift} 
\begin{equation}
\mathbf{L}_{t}\mathbf{=}\frac{J_{H}(1-\cos \omega t)}{J_{H}^{2}+\frac{1}{2}%
\Delta B_{z}^{2}(1+\cos \omega t)}\mathbf{K}^{\dagger }\mathbf{K,}
\label{Eq89+1}
\end{equation}%
where $\omega =\sqrt{J_{H}^{2}+\Delta B_{z}^{2}}$. In the limit of small
magnetic field inhomogeneity, $|\Delta B_{z}/J_{H}|\ll 1$, $\omega
\rightarrow |J_{H}|$ and Eqs. (\ref{Eq88+1}) and (\ref{Eq89+1}) go over into
(\ref{Eq88}) and (\ref{Eq89}), respectively. Eq. (\ref{Eq89+1}) describes
the Lamb energy shift of the triplet state $\left\vert
S=1,M_{S}=0\right\rangle $ due to the coupling between singlet and triplet
states induced by the magnetic field inhomogeneity $\Delta B_{z}$. At the
magnetic field geometry we have chosen there is no coupling between $%
\left\vert S=1,M_{S}=\pm 1\right\rangle $ and $\left\vert
S=1,M_{S}=0\right\rangle $ states.

In Figs. \ref{Fig1}-\ref{Fig3}, we show the results of calculations for the triplet states
population, purity, and the Lamb shift\ energy, respectively, as a function
of time at a fixed interdot separation ($r_{AB}=400\,\mathrm{\mathring{A}}$)
and different $\Delta B_{z}$. For the Heisenberg interaction constant $J_{H}$
we used an asymptotically correct expression \cite{GK03,GP63,HF64} obtained
for hydrogenlike centers in GaAs [note that our $J_{H}=\varepsilon
_{t}-\varepsilon _{s}$ is related to the exchange integral $J$ in Ref.~\onlinecite{GK03}
via $J_{H}=-2J$]. Initially, the system is assumed to be in the $\left\vert
S=1,M_{S}=0\right\rangle $ state. As can be seen from Fig. 1, there is a
re-distribution between singlet and triplet states population due to the
singlet-triplet subspace coupling. At $|\Delta B_{z}/J_{H}|\lesssim 0.1$,
the probability of re-distribution is negligible and the time-evolution is
basically unitary. With increasing $\Delta B_{z}$, this probability
re-distribution is seen to be more pronounced, time-evolution becomes
non-unitary (Fig. 2), and $|a_{t}(t)|^{2}$ can drop to the value $%
J_{H}^{2}/\omega ^{2}$ at $t=\pi n/\omega ,\,n=1,3,\ldots $. Observe that
the non-unitary dynamics reveals repetitions in time and at moments of
maximal (minimal) singlet-triplet states probability re-distribution we find
maximal (minimal) Lamb energy shifts (Fig. 3). Thus, the non-unitary effects
observed are not irreversible and they do not result in a real decoherence
process. We do not have in our two-electron model a real, external and infinite
\textquotedblleft bath\textquotedblright , coupling to which would result in
irreversible decoherence effects in the spin system.  In Figs. 4,5 we
demonstrate the dependence of triplet states population and Lamb energy
shifts on the interdot distance $r_{AB}$ at a fixed $\Delta B_{z}=0.05%
\mathrm{T}$.

\subsection{Spin-orbit interaction in \textit{pure}-spin model}

\label{SO-pure}

In this subsection we estimate the non-unitary effects in the \textit{pure}%
-spin model due to spin-orbit interaction. For simplicity we assume that the
external magnetic field is homogeneous and directed along the $z$-axis, with 
$B_{oz}$ being its $z$-component. Since $\delta \vec{B}_{so}$ is a pure
imaginary field (its components are matrix elements between the real states $%
\tilde{\phi}_{A}$ and $\tilde{\phi}_{B}$ of an odd vector function of the
momentum operator, both in vacuum and in the bulk of semiconductors that
lack inversion symmetry, Dresselhaus fields,\cite{Dres55} as well as in
heterostructure zinc-blendes, Rashba fields,\cite{BR84}) we have%
\begin{equation}
(\delta B_{so}^{\pm })^{\ast }=-\delta B_{so}^{\mp },\qquad \delta
B_{soz}^{\ast }=-\delta B_{soz}.  \label{Eq.91}
\end{equation}

Using these relationships, the singlet-triplet spin-orbit coupling can be
written as%
\begin{equation}
H^{st}(\vec{B}_{so})=\left( 
\begin{array}{ccc}
0 & 0 & 0 \\ 
\frac{1}{2}\delta B_{so}^{+} & -\frac{1}{\sqrt{2}}\delta B_{soz} & -\frac{1}{%
2}\delta B_{so}^{-} \\ 
\frac{1}{2}\delta B_{so}^{+} & -\frac{1}{\sqrt{2}}\delta B_{soz} & -\frac{1}{%
2}\delta B_{so}^{-}%
\end{array}%
\right) .  \label{6.4_1}
\end{equation}

The couplings between double-occupancy, singlet and triplet states are seen
to be the same. We assume that $H^{ss}={\rm diag}(\varepsilon
_{s},\varepsilon _{do},\varepsilon _{do}),$ where $\varepsilon _{s}$ and $%
\varepsilon _{do}$ are the singlet and double-occupancy states energies, and 
$H^{tt}={\rm diag}(\varepsilon _{t+},\varepsilon _{t},\varepsilon _{t-})$,
where $\varepsilon _{t}$ is a triplet state energy and $\varepsilon _{t\pm
}=\varepsilon _{t}\pm B_{0z}$. Within these approximations, the 6-by-6
eigenvalue problem Eq. (\ref{Leq5}) is then reduced to computing the roots
of the biquadratic equation \cite{AS64}%
\begin{equation}
E^{4}+a_{3}E^{3}+a_{2}E^{2}+a_{1}E+a_{0}=0  \label{6.4_2}
\end{equation}%
where%
\begin{eqnarray*}
a_{3} &=&-\sum\limits_{i=1}^{4}\varepsilon _{i},\qquad
a_{2}=\sum\limits_{i\neq j}\varepsilon _{i}\varepsilon
_{j}-\sum\limits_{\alpha =x,y,z}|\delta B_{so\alpha }|^{2}, \\
a_{1} &=&-\sum\limits_{i\neq j\neq k}\varepsilon _{i}\varepsilon
_{j}\varepsilon _{k}+ \\
&&\left( |\delta B_{soz}|^{2}+\frac{1}{2}[|\delta B_{sox}|^{2}+|\delta
B_{soy}|^{2}]\right) (\varepsilon _{2}+\varepsilon _{4})+ \\
&&(|\delta B_{sox}|^{2}+|\delta B_{soy}|^{2})\varepsilon _{3}, \\
a_{0} &=&\prod\limits_{i=1}^{4}\varepsilon _{i}-|\delta
B_{soz}|^{2}\varepsilon _{2}\varepsilon _{4}- \\
&&\frac{1}{2}[|\delta B_{sox}|^{2}+|\delta B_{soy}|^{2}]\varepsilon
_{3}(\varepsilon _{2}+\varepsilon _{4}), \\
\varepsilon _{1} &=&\varepsilon _{s},\qquad \varepsilon _{2}=\varepsilon
_{t+},\qquad \varepsilon _{3}=\varepsilon _{t},\qquad \varepsilon
_{4}=\varepsilon _{t-}.
\end{eqnarray*}

For hydrogenlike centers one can estimate the energies $\varepsilon _{do}$
and $\varepsilon _{t}$ as follows. The ground energy of two well separated
hydrogen atoms is $E_{2\mathrm{H}}\approx -27.2\,\mathrm{eV}$ . Using for
GaAs the scaling factor $K_{\mathrm{GaAs}}=m^{\ast }/\varepsilon ^{2}\approx
4.6\cdot 10^{-4}$ one can estimate $\varepsilon _{t}\approx K_{\mathrm{GaAs}%
}E_{2\mathrm{H}}=-12.6\,\mathrm{meV}$. $\varepsilon _{do}$ is located higher
than $\varepsilon _{t}$ due to mainly interelectron repulsion $\tilde{v}_{%
\mathrm{ee}}(AA;AA)$ so that $\varepsilon _{do}-\varepsilon _{t}=\tilde{v}_{%
\mathrm{ee}}(AA;AA)\approx 12.6\,\mathrm{meV}$.

If $\delta B_{so\alpha }\equiv 0,$ $\alpha =x,y,z$, the roots of Eq. (\ref%
{6.4_2}) $E_{i}$ are equal to $\varepsilon _{i}$, $i=1,\ldots ,4$. The two
other roots are $E_{5}=\varepsilon _{do}$ and $E_{6}=\varepsilon _{s}$. The
corresponding eigenvectors are%
\begin{eqnarray}
\left\vert e_{k}\right\rangle &=&d_{k}\left( 0,1,1,-\frac{\delta B_{so}^{-}}{%
E_{k}-\varepsilon _{2}},\frac{\sqrt{2}\delta B_{soz}}{E_{k}-\varepsilon _{3}}%
,\frac{\delta B_{so}^{+}}{E_{k}-\varepsilon _{4}}\right) ^{T},\qquad  \notag
\\
k &=&1,\ldots ,4,  \notag \\
\left\vert e_{5}\right\rangle &=&\frac{1}{\sqrt{2}}\left(
0,1,-1,0,0,0\right) ^{T},\qquad  \notag \\
\left\vert e_{6}\right\rangle &=&\left( 1,0,0,0,0,0\right) ^{T},
\label{6.4_3}
\end{eqnarray}
\ where%
\begin{eqnarray*}
d_{k} &=&\left( 2+[|\delta B_{sox}|^{2}+|\delta B_{soy}|^{2}]\right. \times
\\
&&\left. \left( \frac{1}{(E_{k}-\varepsilon _{2})^{2}}+\frac{1}{%
(E_{k}-\varepsilon _{4})^{2}}\right) +\frac{2|\delta B_{soz}|^{2}}{%
(E_{k}-\varepsilon _{3})^{2}}\right) ^{-1/2}.
\end{eqnarray*}

Notice that the above formulas are not valid in the degenerate case: $%
B_{0z}=0$ and $\varepsilon _{2}=\varepsilon _{3}=\varepsilon
_{4}=\varepsilon _{t}$. In this case\ the biquadratic Eq. (\ref{6.4_2})
reduces to two quadratic ones, two roots of which are degenerate, $%
E_{1}=E_{2}=\varepsilon _{t}$. Formally, one gets singularities in Eq. (\ref%
{6.4_3}) at $E_{1}=E_{2}=\varepsilon _{t}$. Therefore, the simpler,
degenerate case should be analyzed separately and the corresponding formulas
[not shown here] can be derived.

Let us now find the spin-orbit field%
\begin{equation}
\begin{array}{lll}
\delta \vec{B}_{so} & = & \left\langle \tilde{\phi}_{A}\right\vert \vec{B}%
_{so}(\vec{p})\left\vert \tilde{\phi}_{B}\right\rangle \approx \left\langle
\phi _{A}\right\vert \vec{B}_{so}(\vec{p})\left\vert \phi _{B}\right\rangle
\\ 
& = & \int d\vec{r}\,\phi _{A}(|\vec{r}-\vec{R}|)\vec{B}_{so}(-i\nabla _{%
\vec{r}})\phi _{B}(r) \\ 
& = & \vec{B}_{so}(-i\nabla _{\vec{R}})\int d\vec{r}\,\phi _{A}(|\vec{r}-%
\vec{R}|)\phi _{B}(r) \\ 
& = & \vec{B}_{so}(-i\nabla _{\vec{R}})S(R)%
\end{array}
\label{6.4_4}
\end{equation}%
\ where $\vec{B}_{so}(\vec{p})$ is an odd function of the momentum operator $%
\vec{p}=-i\nabla _{\vec{r}}$ and $\vec{R}=\vec{r}_{AB}$. In particular, in
zinc-blende semiconductors such as GaAs, $\vec{B}_{so}$ is cubic in the
components of $\vec{p}$: \cite{Dres55,DP71}%
\begin{eqnarray}
B_{so\alpha } &=&A_{so}p_{\alpha }(p_{\beta }^{2}-p_{\gamma }^{2}),\qquad 
\notag \\
\alpha ,\beta ,\gamma &=&\{\mathrm{cyclic\,permutations\,of\quad }x,y,z\} 
\notag \\
A_{so} &=&\alpha _{so}\left( m^{\ast }\sqrt{2m^{\ast }E_{g}}\right) ^{-1},
\label{6.4_5}
\end{eqnarray}%
where $m^{\ast }$ is the effective mass of the electron, $E_{g}$ is the band
gap [$m^{\ast }\approx 0.072,$ $E_{g}\approx 1.43\,\mathrm{eV}$ for GaAs], $%
p_{x}$, $p_{y}$, $p_{z}$ are components of the momentum along the cubic axes 
$[100]$, $[010]$, and $[001]$ respectively. The dimensionless coefficient $%
\alpha _{so}=0.07$ for GaAs. From Eqs. (\ref{6.4_4}) and (\ref{6.4_5}) we
obtain%
\begin{eqnarray}
B_{so\alpha } &=&iA_{so}\frac{R_{\alpha }(R_{\beta }^{2}-R_{\gamma }^{2})}{%
R^{3}}\left\{ S^{^{\prime \prime \prime }}(R)-\right.  \notag \\
&&\left. \frac{3}{R}S^{^{\prime \prime }}(R)+\frac{3}{R^{2}}S^{^{\prime
}}(R)\right\}  \label{6.4_6}
\end{eqnarray}

The overlap integral \cite{Sla63} $S(r=R/a_{B})=(1+r+r^{2}/3)\exp (-r)$ for
hydrogenlike centers [$a_{B}\approx 92\,\mathrm{\mathring{A}}$ for GaAs] and
Eq. (\ref{6.4_6}) reduces to%
\begin{eqnarray}
B_{sox} &=&i(0.83\,\mathrm{meV})\sin \theta \cos \varphi (\sin ^{2}\theta
\sin ^{2}\varphi -\cos ^{2}\theta )\times  \notag \\
&&\left( -\frac{1}{3}\right) r^{2}\exp (-r)  \label{6.4_7}
\end{eqnarray}%
where $(R,\theta ,\varphi )$ are spherical coordinates of the vector $\vec{R}
$, with other components being obtained by cyclic interchange of indices.

\begin{figure}[tbp]
\epsfig{
file=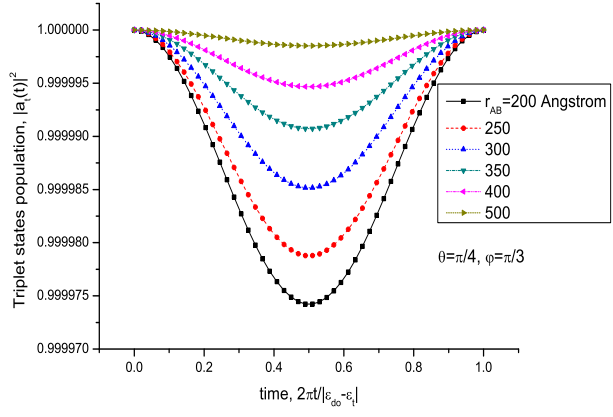,
width=3.5in,height=2.5in} 
\caption{(color online) Triplet states
population dependence on spin-orbit interaction as a function of time, at
different interdot separations $r_{AB}$ and fixed orientation of the
interdot radius-vector, $\vec{r}_{AB}$, ($\protect\theta =\protect\pi /4,%
\protect\varphi =\protect\pi /3$) [see text]. The initial triplet states
population is taken to be equal. The external magnetic field $B_{0z}=0.5$T.}
\label{Fig6}
\end{figure}

\begin{figure}[tbp]
\epsfig{
file=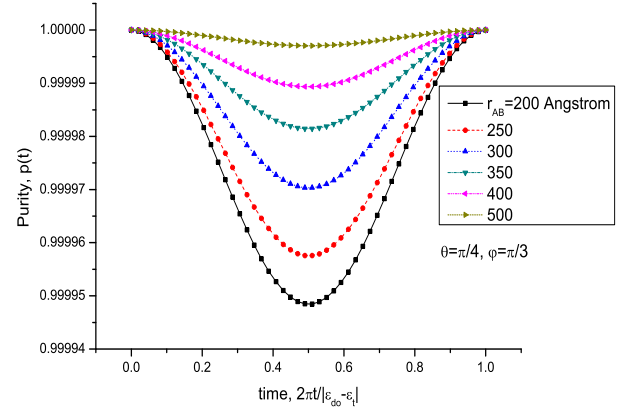,
width=3.5in,height=2.5in} 
\caption{(color online) The purity in the
presence of spin-orbit coupling. All parameters are the same as in Fig. 6.}
\label{Fig7}
\end{figure}

\begin{figure}[tbp]
\epsfig{
file=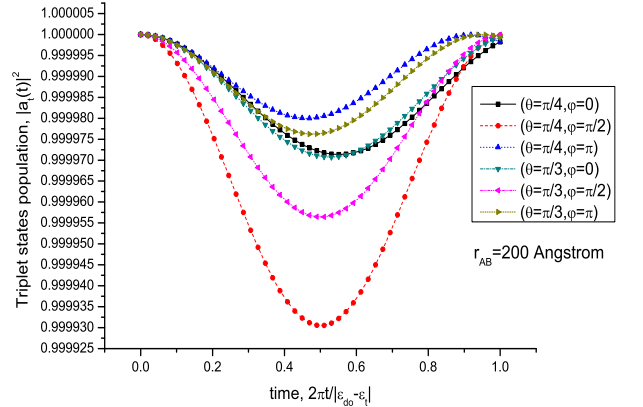,
width=3.5in,height=2.5in} 
\caption{(color online) Effect of
spin-orbit angular anisotropy on triplet states population. All parameters
are the same as in Fig. 6.}
\label{Fig8}
\end{figure}

In Figs.~\ref{Fig6} and \ref{Fig7} we display the time-dependence of the triplet states
population and the purity, which is induced by the spin-orbit interaction,
Eq. (\ref{6.4_7}), at a fixed orientation, ($\theta =\pi /4,\varphi =\pi /3$%
), and different $r_{AB}$ in the range 200-500\AA . Observe that the maximal
re-distribution of singlet-triplet probability occurs at $2\pi
t/|\varepsilon _{do}-\varepsilon _{t}|=0.5$ and the spin-orbit interaction
effect diminishes as $r_{AB}$ increases. The maximal singlet-state
probability achieved at $r_{AB}=200$ $\mathrm{\mathring{A}}$ is seen to be
quite small, $\sim 10^{-5}$. As compared to the non-unitary effects induced
by $\vec{B}$-field inhomogeneity, the spin-orbit effects are on average four
orders of magnitude smaller. The angular dependence of the population of
triplet states on the interdot radius-vector orientation at a fixed $%
r_{AB}=200\,\mathrm{\mathring{A}}$ is illustrated in Fig.~\ref{Fig8}.

\subsection{The $\vec{B}$-field interaction in the \textit{pseudo}-spin model%
}

\label{B-pseudo}

Using Eqs.~(\ref{Eq85}) and (\ref{Eq87}), the effective Hamiltonian matrix
Eq.~(\ref{5.6}) in the basis $\left\{ \Phi _{s1},\Phi _{ti},i=1,2,3\right\} $
can be rewritten as%
\begin{equation}
\mathcal{H}_{\mathrm{eff}}=\left( 
\begin{array}{cccc}
\varepsilon _{s} & \frac{1}{2\sqrt{2}}\Delta B^{+} & -\frac{1}{2}\Delta B_{z}
& -\frac{1}{2\sqrt{2}}\Delta B^{-} \\ 
\frac{1}{2\sqrt{2}}\Delta B^{-} & \varepsilon _{t}+B_{avz} & B_{av}^{-} & 0
\\ 
-\frac{1}{2}\Delta B_{z} & B_{av}^{+} & \varepsilon _{t} & B_{av}^{-} \\ 
-\frac{1}{2\sqrt{2}}\Delta B^{+} & 0 & B_{av}^{+} & \varepsilon _{t}-B_{avz}%
\end{array}%
\right)  \label{6.5_1}
\end{equation}%
where for simplicity we neglected contributions from the double-occupancy
states [the resolvent term in Eq. (\ref{5.2})]. Alternatively, in the 
\textit{pseudo}-spin representation we get%
\begin{eqnarray}
\mathcal{H}_{\mathrm{eff}} &=&\left( \frac{1}{4}\varepsilon _{s}+\frac{3}{4}%
\varepsilon _{s}\right) I+J_{H}\,\vec{s}_{A}\cdot \vec{s}_{B}+  \notag \\
&&\vec{B}_{av}\cdot (\,\vec{s}_{A}+\vec{s}_{B})+\frac{\Delta \vec{B}}{2}%
\cdot (\,\vec{s}_{A}-\vec{s}_{B})  \notag \\
&=&\left( \frac{1}{4}\varepsilon _{s}+\frac{3}{4}\varepsilon _{t}\right)
I+J_{H}\,\vec{s}_{A}\cdot \vec{s}_{B}+  \notag \\
&&\vec{B}_{A}\cdot \vec{s}_{A}+\vec{B}_{B}\cdot \vec{s}_{A},  \label{6.5_2}
\end{eqnarray}%
where $\vec{B}_{A}=\vec{B}_{av}+\Delta \vec{B}/2$ and $\vec{B}_{B}=\vec{B}%
_{av}-\Delta \vec{B}/2$ are the local magnetic fields at sites $A$ and $B$,
respectively. The term $J_{H}\,\vec{s}_{A}\cdot \vec{s}_{B}$ is the familiar Heisenberg interaction. In matrix form, Eq. (\ref{6.5_2}) can be rewritten as%
\begin{eqnarray}
\mathcal{H}_{\mathrm{eff}} &=&\left( 
\begin{array}{cccc}
\varepsilon _{1} & \frac{1}{2}B_{A}^{-} & 0 & 0 \\ 
B_{A}^{+} & \varepsilon _{2} & \frac{1}{2}J_{H} & 0 \\ 
0 & \frac{1}{2}J_{H} & \varepsilon _{3} & \frac{1}{2}B_{B}^{-} \\ 
0 & 0 & \frac{1}{2}B_{B}^{+} & \varepsilon _{4}%
\end{array}%
\right)  \label{6.5_2m} \\
\varepsilon _{1} &=&\varepsilon _{t}+\frac{1}{2}B_{Az},\quad \varepsilon
_{2}=\varepsilon _{t}-\frac{1}{2}(J_{H}+B_{Az}),  \notag \\
\varepsilon _{3} &=&\varepsilon _{t}-\frac{1}{2}(J_{H}-B_{Bz}),\quad
\varepsilon _{4}=\varepsilon _{t}-\frac{1}{2}B_{Bz}  \notag
\end{eqnarray}%
where $B_{A,B}^{\pm }=B_{A,Bx}\pm iB_{A,By}$.

The Hamiltonian Eq. (\ref{6.5_1}) generates a unitary evolution%
\begin{equation}
U_{\mathrm{eff}}(t)=\exp (-it\mathcal{H}_{\mathrm{eff}})  \label{6.5_2+1}
\end{equation}%
in $%
\mathbb{C}
^{4}$. At a fixed set of parameters $\varepsilon _{s},\varepsilon _{t},\vec{B%
}_{A},\vec{B}_{B}$ the propagator $U_{\mathrm{eff}}(t)$ does not provide a
universal set of unitary gates in $%
\mathbb{C}
^{4}$. Any unitary transformation $U\in U(4)$ can be represented as a
product of a phase factor $\exp (i\alpha )$, where $\alpha $ is a real
parameter, and a unitary transformation $U_{S}\in SU(4)$. Any transformation 
$U_{S}$ \ is determined by $M=4^{2}-1=15$ independent real parameters $%
(\theta _{1},\ldots ,\theta _{15})$ so that%
\begin{equation}
U_{S}(\theta _{1},\ldots ,\theta _{M})=\exp \left(
-i\sum\limits_{i=1}^{M}\theta _{i}F_{i}\right) ,  \label{6.5_3}
\end{equation}%
where the set of generators $\{F_{i}\}$ is an orthonormalized traceless,
Hermitian matrix set that forms a Lie algebra $su(4)$ [$F_{i}$'s form a
complete basis in a real $M$-dimensional vector space; they are analogs of
Pauli matrices, $\sigma _{\alpha }$, $\alpha =x,y,z$, in $su(2)$ - see,
e.g., Ref.~\onlinecite{AL87}]. From the\ representation Eq. (\ref{6.5_3}) it follows
immediately that $U_{\mathrm{eff},S}(t)=\exp \left( -it\left( \frac{1}{4}%
\varepsilon _{s}+\frac{3}{4}\varepsilon _{t}\right) \right) U_{\mathrm{eff}%
}(t)$ cannot match an arbitrary $U_{S}$, because the number of independent parameters
in Eq. (\ref{6.5_1}) is at most $8$ -- fewer than the number of $\theta _{i}$%
's. This can also be understood from the fact that the form of the
Hamiltonian matrix, Eq. (\ref{6.5_1}), is not generic. In particular, the
matrix is sparse, i.e., the entries $(2,4)$ and $(4,2)$ are zeros.

However, compositions of unitary transformations Eq. (\ref{6.5_2+1}) taken
at different sets of parameters can provide a universal set of unitary gates
in $%
\mathbb{C}
^{4}$. A well-known example of universal gates is provided by the Heisenberg
interaction (at $J_{H}\neq 0$) with single-spin addressing (at $\Delta \vec{B%
}\neq 0$).\cite{Loss:98}

From Eq.~(\ref{6.5_1}) it follows that a necessary and sufficient condition
for obtaining a universal set of gates on two spins is to have an inhomogeneity in the
magnetic field $\Delta \vec{B}\neq 0$ [the source of inhomogeneity can be
different, it can be either strongly localized magnetic fields or $g$-factor
engineering] and the Heisenberg interaction, $J_{H}\neq 0$. The reason is
that when $\Delta \vec{B}=0$, the Hamiltonian Eq. (\ref{6.5_1}) and the
corresponding unitary transformations take a block-diagonal form, with
singlet-triplet entries being zeros, while when $J_{H}=0$, \ the Hamiltonian
form (\ref{6.5_2m}) will have zero off-diagonal block-matrices. Clearly,
even a composition of such unitary transformations taken at different sets
of parameters, either $\varepsilon _{s},\varepsilon _{t},\vec{B}_{A}=\vec{B}%
_{B}$ or $\varepsilon _{s}=\varepsilon _{t},\vec{B}_{A},\vec{B}_{B}$, will
be in a block-diagonal form and it cannot reproduce an arbitrary unitary
transformation. Note that when one allows for encoding a qubit into three
or more spins, the Heisenberg interaction alone is universal in the
pseudo-spin model,\cite{Bacon:99a,Kempe:00} and Heisenberg along with
an inhomogneous magnetic field is universal for an encoding of a
single qubit into pair of spins.\cite{LidarWu:01}

Moreover, it should be noted that in the homogeneous magnetic field case
unitary transformations restricted to the triplet subspace will not provide
a universal set of gates. To prove this statement, let us consider a
composition of two unitary transformations in the triplet subspace:%
\begin{equation}
\begin{array}{l}
\exp (-it_{1}H^{tt}(\vec{B}_{1}))\exp (-it_{2}H^{tt}(\vec{B}_{2}))= \\ 
\exp (-i(t_{1}+t_{2})H_{\mathrm{eff}}^{tt})= \\ 
\exp \left( -i(t_{1}+t_{2})\varepsilon _{t}-it_{1}H^{tt}(\vec{B}%
_{1})-it_{2}H^{tt}(\vec{B}_{2})-\right. \\ 
\left. \frac{t_{1}t_{2}}{2}\left[ H^{tt}(\vec{B}_{1}),H^{tt}(\vec{B}_{2})%
\right] +\cdots \right) ,%
\end{array}
\label{6.5_4}
\end{equation}%
where on the right-hand-side we used the Campbell-Hausdorff
formula.\cite{Cor90}
From Eq. (\ref{Eq85}) one obtains%
\begin{equation}
\left[ H^{tt}(\vec{B}_{1}),H^{tt}(\vec{B}_{2})\right] =i\left( 
\begin{array}{ccc}
2\mathcal{B}_{z} & \mathcal{B}^{-} & 0 \\ 
\mathcal{B}^{+} & 0 & \mathcal{B}^{-} \\ 
0 & \mathcal{B}^{+} & -2\mathcal{B}_{z}%
\end{array}%
\right) ,  \label{6.5_5}
\end{equation}%
where $\mathcal{\vec{B}=}[\vec{B}_{1}\times \vec{B}_{2}]$ and $\mathcal{B}%
^{\pm }=\mathcal{\vec{B}}_{x}\pm i\mathcal{\vec{B}}_{y}$. Since the
higher-order terms in the Campbell-Hausdorff formula consist of nested
commutators between $H^{tt}(\vec{B}_{1})$ and $H^{tt}(\vec{B}_{2})$, we find
that the effective Hamiltonian $H_{\mathrm{eff}}^{tt}$ corresponding to the
product of two unitary transformations will still have a sparse form, with
the $(1,3)$ and $(3,1)$ entries being zeros.

\subsection{Is it possible to obtain a universal set of gates in the \emph{%
pure}-spin model?}

\label{universal}

Simultaneously, we have just proved that in the \emph{pure}-spin model, in
the case of a homogeneous magnetic field, unitary transformations in the
triplet subspace will not provide a universal set of gates. On the other
hand, at $\Delta \vec{B}\neq 0$, we have already shown that the evolution of
the spin-density matrix is non-unitary. Let us assume that we have a
non-unitary gate $\mathcal{L}_{1}$ so that $\rho (t_{1})=\mathcal{L}%
_{1}(t_{1})\rho (0)$. How could one define a composition of two non-unitary
gates, $\mathcal{L}_{2}\mathcal{L}_{1}$? In order to do this unambiguously, $%
\mathcal{L}_{2}$ should obey a compatibility condition with the initial
state [because a non-unitary $\mathcal{L}$-gate is not totally independent
of the initial state, it includes some sort of correlation information
encoded in the initial state], that is a correlation $R_{m}(t_{1})$
established between $a_{s}(t_{1})$ and $a_{t}(t_{1})$ amplitudes at $t=t_{1}$
should be included in in the definition of the corresponding dynamics
generator operators in $\mathcal{L}_{2}$. Eq. (\ref{Leq4}), where the
left-hand-side and $H^{ts}$ should be replaced by $R_{m}(t_{1})$ and the
identity matrix, respectively, provides a relationship between $R_{m}(t_{1})$
and $R_{m}(0)$. If the correlation between the amplitudes at $t=0$ and $%
t=t_{1}$ is the same, $R_{m}(t_{1})=$ $R_{m}(0)$, then we obviously have $%
\mathcal{L}_{1}=\mathcal{L}_{2}=\mathcal{L}$ and $\mathcal{L}_{2}(t_{2})%
\mathcal{L}_{1}(t_{1})=\mathcal{L(}t_{1}+t_{2})$.

In the total Hilbert space, the state is defined by 11 real parameters.
While in the reduced description, the spin-density matrix is defined by 5
real parameters (for more on the spin density matrix parametrization in
terms of $a$'s, see Section \ref{sec:conc}). Fixing a correlation in the
initial state, $a_{s}(0)=R_{m}(0)a_{t}(0)$, we have 3 complex equations
between the amplitudes $a_{s}(0)$ and $a_{t}(0)$, which define a 5D real
manifold embedded into the total Hilbert space. Using these equations we can
separate 6 extra real degrees of freedom that we have in the total state
description from those in the spin-density matrix description. However,
these extra degrees of freedom are not eliminated in the spin-density
description, they are included in the form of correlation matrix $R_{\alpha
}(0)$, $\alpha =\{s,t,m\}$. It was shown in part I that Eq. (\ref{Leq1})
provides an \emph{exact} description of quantum evolution, in the
spin-density space. Therefore, as long as we have a universal set of unitary
gates in the total Hilbert space, this set of gates will be translated into
the corresponding universal set of non-unitary gates generated by Eq. (\ref%
{Leq1}) because no information is lost in our \textquotedblleft
reduced\textquotedblright\ spin-density matrix description.

\subsection{Spin-orbit interaction in the \textit{pseudo}-spin model}

\label{SO-pseudo}

Let us consider spin-orbit effects, which are proportional to $\delta B_{so}$%
, in the \textit{pseudo}-spin model. From Eq. (\ref{5.6}) we obtain%
\begin{equation}
\mathcal{H}_{so}(\delta \vec{B}_{so})=\sum_{kk^{\prime
}=2,3}H_{1k}^{ss}\left( \frac{1}{\bar{E}I-H^{ss}}\right) _{kk^{\prime
}}^{-1}F_{k^{\prime }}(\delta \vec{B}_{so}),  \label{Eq.90}
\end{equation}%
where 
\begin{equation}
F_{k}(\delta \vec{B}_{so})=\sum_{i=1}^{3}\left( H_{ki}^{st}(\delta \vec{B}%
_{so})\mathbf{K}_{i}+(H_{ki}^{st}(\delta \vec{B}_{so}))^{\ast }\mathbf{K}%
_{i}^{\dagger }\right)  \label{Eq.90.1}
\end{equation}

It follows from Eq. (\ref{6.4_1}) that $H_{2i}^{st}=H_{3i}^{st}$, $%
F_{2}=F_{3}$ and Eq.~(\ref{Eq.90.1}) can be reduced to%
\begin{equation}
F_{2,3}(\delta \vec{B}_{so})=F(\delta \vec{B}_{so})=i\sqrt{2}\delta \vec{B}%
_{so}\left[ \vec{s}_{A}\times \vec{s}_{B}\right]  \label{Eq.92}
\end{equation}%
Then, Eq. (\ref{Eq.90}) can be rewritten as%
\begin{equation}
\mathcal{H}_{so}(\delta \vec{B}_{so})=\mathcal{A}_{do}F(\delta \vec{B}_{so})
\label{Eq.93}
\end{equation}%
where the coefficient $\mathcal{A}_{do}$ is proportional to the ratios of
the amplitudes of double occupancy transitions ($H_{12}^{ss}$ and $%
H_{13}^{ss}$) and the energies of interelectron interaction ($H_{22}^{ss}-%
\bar{E}$ and $H_{33}^{ss}-\bar{E}$) in doubly occupied QDs:%
\begin{eqnarray}
\mathcal{A}_{do} &=&\sum_{kk^{\prime }=2,3}H_{1k}^{ss}\left( \frac{1}{\bar{E}%
I-H^{ss}}\right) _{kk^{\prime }}  \notag \\
&=&\frac{1}{\Delta }\left[ H_{12}^{ss}(\bar{E}-H_{33}^{ss}-H_{32}^{ss})+%
\right.  \notag \\
&&\left. H_{12}^{ss}(\bar{E}-H_{22}^{ss}-H_{23}^{ss})\right]  \notag \\
&\approx &\frac{H_{12}^{ss}}{\bar{E}-H_{22}^{ss}}+\frac{H_{13}^{ss}}{\bar{E}%
-H_{33}^{ss}}.  \label{Eq.93.1}
\end{eqnarray}%
Here, the determinant is%
\begin{equation}
\Delta =(\bar{E}-H_{22}^{ss})(\bar{E}-H_{33}^{ss})-|H_{23}^{ss}|^{2}.
\label{Eq.93.2}
\end{equation}

The effective spin-orbit interaction Hamiltonian Eq. (\ref{Eq.93}) is
different from the corresponding one obtained by Kavokin.\cite{Kav01} In
our derivation the double-occupancy states are essential, whereas in Ref.~\onlinecite
{Kav01} these states are totally neglected. We showed above that neglecting
double-occupancy states results in zero spin-orbit coupling. The very
physical picture put forward in Ref.~\onlinecite{Kav01} to support the derivation was
based on the assumption that when one of the two electrons localized at
centers $A$ or $B$ tunnels to the adjacent center (say, from $A$ to $B$), it
experiences the influence of the spin-orbit field resulting from the
under-barrier motion of the electron. Neglecting double occupancy states
means that the second electron should simultaneously tunnel from $B$ to $A$
so that the two electrons can never be found in the same QD. Indeed,
in Ref.~\onlinecite{Kav01} this simultaneous two-electron transition is
described by the 
product of two matrix elements: the overlap $S$ and $(\delta \vec{B}%
_{so})_{\alpha }$, $\alpha =x,y,z$ [$\mathcal{H}_{so}\sim S(\delta \vec{B}%
_{so}\cdot \left[ \vec{s}_{A}\times \vec{s}_{B}\right] )$, our $\delta \vec{B%
}_{so}$ is related to Kavokin's $\mathbf{b}$-field via $\delta \vec{B}%
_{so}=-i\mathbf{b}$, and the overlap via $S=\Omega $]. With the
orthogonalized molecular-type, two-center orbitals such a one step
two-electron transition gives a zero contribution since the spin-orbit
interaction is a one-electron operator and the overlap $\tilde{S}%
=\left\langle \tilde{\phi}_{A}\right\vert \left. \tilde{\phi}%
_{B}\right\rangle =0$. Eq. (\ref{Eq.90}) describes the two-step mechanism:
in the first step, the two-electron system makes a transition from the
singly-occupied state $\Phi _{s1}$ to the intermediate, double-occupancy
states $\Phi _{s2}$, $\Phi _{s3}$ due to the inter-electron interaction ($%
H_{1k}^{ss},$ $k=1,2$ terms). Then in the second step, as a result of the
spin-orbit interaction, the system makes transitions from $\Phi _{s2}$, $%
\Phi _{s3}$ to $\Phi _{ti}$ triplet states (the $H_{ki}^{st}$ terms).

Let us find an estimate for%
\begin{eqnarray}
\mathcal{A}_{do}(R) &\approx &-\frac{2\sqrt{2}}{|\varepsilon
_{do}-\varepsilon _{t}|}\iint d\vec{r}_{1}d\vec{r}_{2}\,\phi
^{2}(r_{1})\times  \notag \\
&&\phi (|\vec{r}_{2}-\vec{R}|)\frac{1}{\varepsilon |\vec{r}_{2}-\vec{r}_{1}|}%
\phi (r_{2}),  \label{Eq.93.3}
\end{eqnarray}%
where the hydrogenlike orbital $\phi (r)=(\pi a_{B}^{3})^{-1/2}\exp
(-r/a_{B})$. Since electron 1 in Eq. (\ref{Eq.93.3}) is localized around the
effective Bohr radius $a_{B}$, one can approximate%
\begin{equation}
\frac{1}{|\vec{r}_{2}-\vec{r}_{1}|}\approx \frac{1}{|\vec{r}_{2}-a_{B}\vec{r}%
_{1}/r_{1}|}.  \label{Eq.93.4}
\end{equation}%
Then, the remaining integrals can be calculated exactly and we obtain%
\begin{equation}
\begin{array}{ccc}
\mathcal{A}_{do}(r=R/a_{B}) & \approx & -\frac{4\sqrt{2}}{er}\left\{
4eF_{1}(r)\right. - \\ 
&  & 2\left[ F_{1}(r+1)+F_{1}(r-1)\right] - \\ 
&  & \left[ F_{2}(r+1)-F_{2}(r-1)\right] + \\ 
&  & \left. 2\left[ F_{0}(r+1)-F_{0}(r-1)\right] \right\}%
\end{array}
\label{Eq.93.5}
\end{equation}%
where%
\begin{equation}
\begin{array}{lll}
F_{0}(x) & = & \frac{{\rm sign}(x)}{48}\left[ 15(1+|x|)+6|x|^{2}+|x|^{3}%
\right] \exp (-|x|), \\ 
F_{1}(x) & = & \frac{1}{48}\left[ 3|x|(1+|x|)+|x|^{3}\right] \exp (-|x|), \\ 
F_{2}(x) & = & \frac{{\rm sign}(x)}{2}(1+|x|)\exp (-|x|).%
\end{array}
\label{Eq.93.6}
\end{equation}%
Here, we used the same estimate for $\varepsilon _{do}-\varepsilon _{t}$ as
in Section \ref{SO-pure}.

\begin{figure}[tbp]
\epsfig{
file=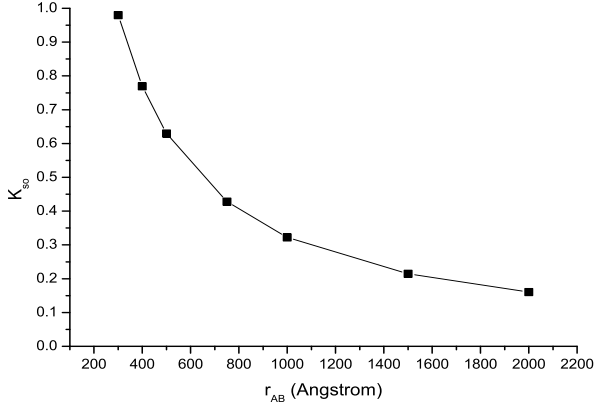,
width=3.5in,height=2.5in} 
\caption{Spin-orbit interaction
reduction coefficient Eq. (\protect\ref{Eq.93.7}) as a function of interdot
distance for GaAs.}
\label{Fig9}
\end{figure}

In order to compare our calculations to Kavokin's result for GaAs, we have
plotted in Fig.~\ref{Fig9} the spin-orbit interaction reduction coefficient%
\begin{equation}
K_{so}=\frac{|\mathcal{A}_{do}|}{\sqrt{2}S},  \label{Eq.93.7}
\end{equation}%
which is exactly the ratio of our and Kavokin's estimates, as a function of
interdot distance. As one can see, $K_{so}$ decreases from 0.98 to 0.46 in
the range $r_{AB}\sim 300-700\mathrm{\mathring{A}}$. Interestingly, our
results qualitatively agree with the results of Ref.~\onlinecite{GK03}, which
obtained in the region of interest [$r_{AB}\sim (3-7)a_{B}$] a reduction of
about one-half relative to that of Refs.~\onlinecite{Kav01,DK02}. According to
Ref.~\onlinecite{GK03}, $\mathcal{H}_{so}\sim 2J\gamma _{\mathrm{GK}}$, where $J$
is the exchange integral calculated using the medium hyperplane method \cite%
{GP63,HF64} and $\gamma _{\mathrm{GK}}$ is an angle of spin rotation due to
spin-orbit interaction introduced by Gor'kov and Krotkov [$\gamma _{\mathrm{%
GK}}\approx \frac{1}{2}\gamma _{\mathrm{K}}$, $\gamma _{\mathrm{K}}$ being
the corresponding angle of spin rotation introduced by Kavokin]. Note that
Kavokin's $\gamma _{\mathrm{K}}$ and $J$ are not independent parameters. In 
Ref.~\onlinecite{Kav01}, $\gamma _{\mathrm{K}}$ was defined as $\Omega b/J$ so that
their product $2J\gamma _{\mathrm{K}}=2\Omega b$ does not depend explicitly
on $J$.

\section{Electrons on Liquid Helium}

\label{sec:e-on-LHe}

Recently, Lyon suggested that the spin of electrons floating on the
surface of liquid helium (LHe) will make an excellent qubit.\cite{Lyon06} Lyon's
proposal, instead of using the spatial part of the electron wavefunction as a
qubit as in the charge-based proposal,\cite{PD99,DP00,DPS03,DPS04} takes
advantage of the smaller vulnerability of the electron's spin to external magnetic
perturbations and, as a consequence, a longer spin-coherence time. It also
has the important advantage over semiconductor spin-based proposals
(as first pointed out in the charge-based
proposal\cite{PD99,DP00,DPS03,DPS04}) that, with the
electrons residing in the vacuum, several important sources of spin
decoherence are eliminated so that the environment effects are highly
suppressed (the spin-coherence time is estimated\cite{Lyon06} to be $T_{2}>100$s).

The geometry of the system with the electrons trapped at the LHe-vacuum
interface (see Fig.~2 in Ref.~\onlinecite{Lyon06}), is conceptually similar to
that of
semiconductor heterostructures. The two electrons are trapped at sites $%
\vec{r}_{A}$ and $\vec{r}_{B}$ ($\vec{r}_{A,B}$ are radius-vectors of the
centers of quantum dots in the $z=0$ plane) by the two attractive
centers created by two charged spherical electrodes located below the
LHe surface at distance $h$ and separated by the interdot distance $r_{AB}=|%
\vec{r}_{A}-\vec{r}_{B}|$:%
\begin{eqnarray}
V_{\rm tr}(z,\vec{r}) &=&-\frac{\Lambda }{z}+V_{A}(z,\vec{r})+V_{B}(z,\vec{r})
\label{4.1} \\
V_{A,B}(z,\vec{r}) &=&-\frac{Q_{A,B}}{\sqrt{\left( z+h\right) ^{2}+\left( 
\vec{r}-\vec{r}_{A,B}\right) ^{2}}} ,  \label{4.2}
\end{eqnarray}%
where the first potential, $-\Lambda /z$, is due to attraction to the image
charge induced by an electron in the LHe [$\Lambda =(\varepsilon
-1)/(4(\varepsilon +1))\approx 7\cdot 10^{-3}$, with $\varepsilon \approx
1.057$ being the dielectric constant of helium]. For purposes of interaction
control, the $Q_{A,B}$ charges on the electrodes can be made variable in time.
The electrons are prevented from penetrating into helium by a high potential
barrier ($\sim 1$ eV) at the helium surface, so that formally one can put $%
V_{\rm tr}=\infty $ at $z<0$. The in-plane and out-of-plane motion of electrons
in the potential of Eq.~(\ref{4.1}) is in general non-separable. However, near
the electrode's position $\vec{r}_{A,B}$ the potential of Eq.~(\ref{4.2}) is
approximately separable in the $z$ and $\vec{r}$ coordinates 
\begin{equation}
V_{A,B}(z,\vec{r})\approx -\frac{Q_{A,B}}{h}+\mathcal{E}_{\perp A,B}\,z+%
\frac{1}{2}\omega _{A,B}^{2}\left( \vec{r}-\vec{r}_{A,B}\right) ^{2} ,
\label{4.3}
\end{equation}%
where it is assumed that $z$ and $|\vec{r}-\vec{r}_{A,B}|\ll h$ and $%
\mathcal{E}_{\perp A,B}=Q_{A,B}/h^{2}$, $\omega _{A,B}=\left(
Q_{A,B}/h^{3}\right) ^{1/2}$. In the separable approximation of
Eq. (\ref{4.3}), the
electron's motion in the $z$-direction [$z>0$] is described by a 1D-Coulomb
potential perturbed by a small Stark interaction and the in-plane motion by a
2D-oscillatory potential. We assume that the out-of-plane motion in
the $z$%
-direction is ``frozen'' in the ground state of a 1D-Coulomb potential%
\begin{equation}
\varphi _{0}^{C}(z)=2\sqrt{\Lambda }(\Lambda z)\exp (-\Lambda z) , \label{4.4}
\end{equation}%
and the in-plane motion, in a superposition of the in-plane confining
oscillatory potential and, possibly, a perpendicular magnetic field, is
described by the Fock-Darwin (FD) states of Eq.~(\ref{Eq64}). Then, the
calculation of $h_{ij}$ and $v_{ee}(ij;kl)$ in
Eqs. (\ref{Eq72})-(\ref{Eq74}) in the chosen basis set is reduced to
the one-dimensional integrals 

\begin{eqnarray}
h_{AB} &=&\left[ -\frac{\Lambda }{2}+\frac{1}{4l_{A}^{2}}-2\Lambda
Q_{A}g_{c}(\alpha ,\beta ,\lambda _{A})-\right.  \notag \\
&&\left. 2\Lambda Q_{B}g_{c}(\alpha ,\beta ,\lambda _{B})\right] S_{AB}, 
\notag \\
h_{AA} &=&-\frac{\Lambda }{2}+\frac{1}{4l_{A}^{2}}-2\Lambda
Q_{A}g_{c}(\alpha ,\beta _{A},0)-  \notag \\
&&2\Lambda Q_{B}g_{c}(\alpha ,\beta _{A},2\Lambda r_{AB}),  \notag \\
h_{BB} &=&-\frac{\Lambda }{2}+\frac{1}{4l_{B}^{2}}-2\Lambda
Q_{A}g_{c}(\alpha ,\beta _{B},2\Lambda r_{AB})-  \notag \\
&&2\Lambda Q_{B}g_{c}(\alpha ,\beta _{B},0),  \notag \\
g_{c}(\alpha ,\beta ,\lambda ) &=&\int_{0}^{\infty }dx\,J_{0}(\lambda
x)\times  \notag \\
&&\exp \left( -\alpha x-\beta x^{2}\right) /(x+1)^{3},  \label{4.5} \\
\alpha &=&2\Lambda h,\quad \beta =\frac{\left( 2\Lambda l_{A}l_{B}\right)
^{2}}{l_{A}^{2}+l_{B}^{2}},\quad  \notag \\
\beta _{A,B} &=&2(\Lambda l_{A,B})^{2},\quad \lambda _{A,B}=\frac{2\Lambda
l_{B,A}^{2}}{l_{A}^{2}+l_{B}^{2}}r_{AB},  \notag
\end{eqnarray}

\begin{eqnarray}
v_{ee}(AB;CD) &=&N_{ee}g_{ee}(a,b),  \notag \\
N_{ee} &=&\frac{\Lambda l_{A}l_{B}l_{C}l_{D}}{%
(l_{A}^{2}+l_{C}^{2})(l_{B}^{2}+l_{D}^{2})}\times  \notag \\
&&\exp \left( -\frac{r_{AC}^{2}}{4(l_{A}^{2}+l_{C}^{2})}-\frac{r_{BD}^{2}}{%
4(l_{B}^{2}+l_{D}^{2})}\right) ,  \notag \\
g_{ee}(a,b) &=&\int_{0}^{\infty }dx\,J_{0}(bx)\times  \notag \\
&&(3x^{2}+9x+8)(x+1)^{-3},  \label{4.6} \\
a &=&4\Lambda ^{2}\left( \frac{l_{A}^{2}l_{C}^{2}}{l_{A}^{2}+l_{C}^{2}}+%
\frac{l_{B}^{2}l_{D}^{2}}{l_{B}^{2}+l_{D}^{2}}\right) ,\quad  \notag \\
b &=&2\Lambda \left\vert \frac{l_{C}^{2}\vec{r}_{A}+l_{A}^{2}\vec{r}_{C}}{%
l_{A}^{2}+l_{C}^{2}}-\frac{l_{D}^{2}\vec{r}_{B}+l_{B}^{2}\vec{r}_{D}}{%
l_{B}^{2}+l_{D}^{2}}\right\vert , \notag
\end{eqnarray}%
where $i=A,B,C,D$ in the two-electron matrix elements denotes orbitals with
the effective lengths $l_{i}$ localized at $\vec{r}_{i}$, and $J_{0}(x)$
is the zeroth order Bessel function. From Eq.~(\ref{4.6}) one can obtain
the following expression for the Heisenberg interaction constant%
\begin{eqnarray}
J_{H} &=&-1.36\cdot 10^{4}\Lambda g_{ee}(a,0)S^{2}\text{\textrm{\ meV}}
\label{4.7} \\
a &=&\frac{8\Lambda ^{2}l_{A}^{2}l_{B}^{2}}{l_{A}^{2}+l_{B}^{2}}.  \notag
\end{eqnarray}%
Note that $J_{H}$ is proportional to the square of the overlap matrix
element $S$, Eq.~(\ref{Eq71}), and the integral $g_{ee}(a,0)$ does not
depend on the interdot distance $r_{AB}$. As a rough estimate, one can
approximate the rational function in the integral $g_{ee}(a,0)$ by a
constant 8, as a result obtaining $g_{ee}(a,0)\approx 4\sqrt{\pi /a}$.

\begin{figure}[tbp]
\epsfig{
file=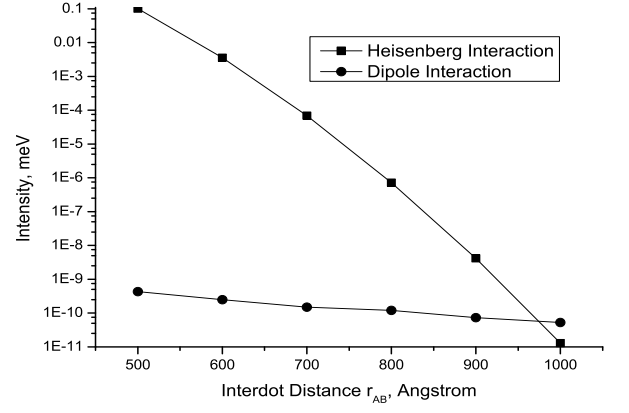,
width=3.5in,height=2.5in} 
\caption{The Heisenberg and
dipole-dipole spin interaction magnitudes as a function of interdot
distance. The distance from the electrodes to the LHe surface $h=800\,%
\mathring{A}$ and the charges on the electrodes $Q_{A}=Q_{B}=1\,\mathrm{a.u.}
$ The magnetic field $\vec{B}_{0}=0$.}
\label{Fig10}
\end{figure}

\begin{figure}[tbp]
\epsfig{
file=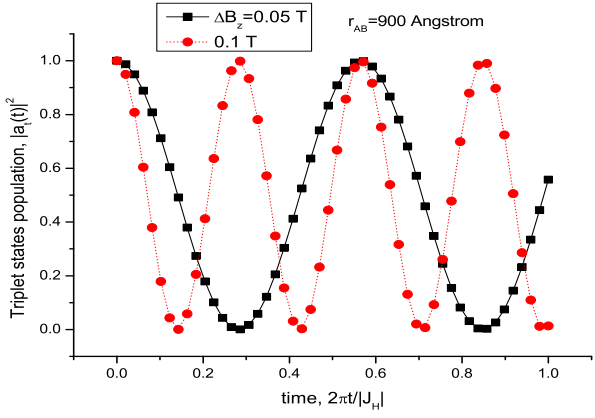,
width=3.5in,height=2.5in} 
\caption{(color online) The triplet
states population of electrons trapped in QDs on a LHe surface, as a
function of time at different magnetic field differences $\Delta B_{z}=0.05$
and 0.1 T. Initially the system is in the triplet state $\left\vert
S=1,M_{S}=0\right\rangle $. The distance between QDs is 900 \AA . All other
parameters are the same as in Fig.~\ref{Fig10}.}
\label{Fig11}
\end{figure}

\begin{figure}[tbp]
\epsfig{
file=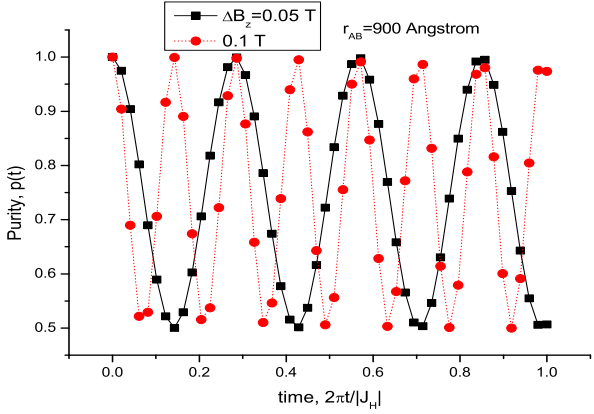,
width=3.5in,height=2.5in} 
\caption{(color online) The purity $p(t)$ for the same parameters as in Fig.~\ref{Fig11}.}
\label{Fig12}
\end{figure}

Figure~\ref{Fig10} shows the magnitudes of the Heisenberg and dipole-dipole spin
interaction, $|J_{H}|$ and $J_{\mathrm{dip}}$ from Eqs.~(\ref{4.7})
and (\ref{Eq84}) respectively, as a function of interdot distance. As
estimated, the magnitude of the Heisenberg interaction is comparable to the weak
dipole-dipole interaction at $r_{AB}\simeq 1000\,\mathrm{\mathring{A}}$.
However, we remark that the strong dependence of $S\sim \exp (-\alpha
r_{AB}^{2})$ (quadratic in the interdot
distance) is due to the
quadratic dependence on coordinates in the exponent of the corresponding
oscillatory wavefunctions. Asymptotically, the confining potential
Eq.~(\ref{4.2}) behaves as a 2D Coulomb potential, so that one should
expect a milder coordinate dependence, $S\sim \exp (-\alpha r_{AB})$, at large
distances (assuming $B_{0}=0$) and the rough estimate of Eq.~(\ref{4.7})
provides a lower bound for the Heisenberg interaction strength.

Lyon suggested\cite{Lyon06} using, instead of the exchange
interaction, the magnetic
dipole-dipole interaction between the spins in order to implement two-qubit
gates, motivating this by the strong sensitivity of the exchange coupling to the
parameters of the system and, hence, the corresponding difficulties with attempting to
control this interaction. Our analysis confirms this, though, of course it is not easy to control the
dipole-dipole interaction either: Eq.~(\ref{Eq84}) shows that this interaction
depends on only one controllable parameter, the interdot distance (the
$g$-factor is a constant in vacuum).

Similarly to Figs.~\ref{Fig1} and \ref{Fig2}, Figs. \ref{Fig11} and \ref{Fig12} demonstrate non-unitary
effects in the \emph{pure}-spin model, due to magnetic field
inhomogeneity in the electrons-on-LHe system. The interdot distance
shown is $r_{AB}=900\,\mathrm{\mathring{A}}$. At this distance
the Heisenberg interaction still prevails over the dipole interaction by at least an
order of magnitude. Again, the pattern seen in the singlet-triplet states
population redistribution is clearly oscillatory.

\section{Discussion and Conclusion}

\label{sec:conc}

We have performed a comparative study of \emph{pure}- and \emph{pseudo}-spin
dynamics for a system of two interacting electrons trapped in two QDs. We
have shown that when there is negligible coupling between the spin and
orbital degrees of freedom, which is the case of near $\vec{B}$-field
homogeneity and negligible spin-orbit interaction, the system spin dynamics
is unitary in both \emph{pure}- and \emph{pseudo}-spin models and is
governed by the Zeeman interaction Hamiltonian of the total spin $\vec{S}$ ($%
S=1$) with the magnetic field $\vec{B}_{av}$. The singlet and triplet states
are totally decoupled; the total spin is conserved. The spin system Hilbert
space can be decomposed into two independent, singlet and triplet subspaces,
the singlet spin states being magnetically inactive ($S=0$). Thus, the
two-electron spin system restricted to the triplet subspace physically
embodies a qutrit. The Heisenberg interaction operates differently in \emph{%
pure}- and \emph{pseudo}-spin models. If for simplicity we neglect
double-occupancy states, the \emph{pseudo}-spin state is totally defined by
four complex amplitudes: $\left\{ a_{s1}(0),a_{ti}(0),\,i=1,2,3\right\} $ in
the basis $\{\Phi _{s1},\Phi _{ti}\}$; so that the Heisenberg interaction
results in a phase unitary transformation: $\{a_{s1}(t)=\exp (-i\varepsilon
_{s}t)a_{s1}(0),$ $a_{ti}(t)=\exp (-i\varepsilon _{t}t)a_{ti}(0)\}$. Since
the spin-density matrix $\rho (t)$ is a bilinear combination of $a$'s: $\rho
_{tij}(t)=a_{ti}(t)a_{tj}^{\ast }(t)$, $\rho _{s}(t)=1-\sum_{i}\rho
_{tii}(t) $, the $\rho $-state will not by affected by the unitary
transformation induced by the Heisenberg interaction.

We have also shown that unitary quantum gates realized in both spin models
do not provide a universal set of gates under the condition $\Delta \vec{B}%
=0 $. In order to obtain a universal set of gates, there should be both
non-zero coupling between singlet and triplet states ($\Delta \vec{B}\neq 0$%
) and non-zero Heisenberg interaction ($J_{H}\neq 0$). Although at $\Delta 
\vec{B}\neq 0$ \emph{pure}-spin dynamics becomes non-unitary, one can
establish a relationship between unitary gates in \emph{pseudo}-spin and the
corresponding non-unitary gates in \emph{pure}-spin dynamics so that a
universal set of quantum gates constructed within the \emph{pseudo}-spin
model will generate a universal set of non-unitary gates in \emph{pure}-spin
dynamics.

To demonstrate the non-unitary effects, which are proportional to the square
of the magnetic field inhomogeneity, in \emph{pure}-spin dynamics, we have
calculated how singlet and triplet states populations, as well as the purity
and the Lamb energy shift, are affected by $\Delta \vec{B}\neq 0$ and
spin-orbit interaction in $n$-doped GaAs semiconductors. These effects are
found to be strongly dependent on the ratio of $\vec{B}$-field inhomogeneity
and the Heisenberg interaction constant, $|\Delta B_{z}/J_{H}|$. For
example, the singlet-triplet states population re-distribution is maximal at 
$t=\pi n/\omega $, where $\omega =\sqrt{J_{H}^{2}+\Delta B_{z}^{2}}$ and $%
n=1,3,\ldots $, and the singlet state population can achieve the value $%
\Delta B_{z}^{2}/\omega ^{2}$. Thus, we can conclude that the Heisenberg
interaction, characterized by the interaction constant $J_{H},$ plays an
essential role in producing non-unitary effects in \emph{pure}-spin
dynamics. Spin-orbit interaction effects are found to be roughly four orders
of magnitude smaller as compared to those caused by $\vec{B}$-field
inhomogeneity.

As shown in Figs.~\ref{Fig1}-\ref{Fig8}, there are clear oscillations in the \emph{pure}%
-spin dynamics and the non-unitary behavior of the spin-density matrix does
not show the decaying pattern characteristic of a real decoherence process. This
should be expected since the ``bath'' -- the electron orbitals -- in our spin model
is not a real stochastic or infinite external bath, an interaction with which may result in
irreversible decoherence. In essence, the spin dynamics is embedded
in space and our bath is too small. The coordinate Hilbert space in the
two-orbital ground state approximation adopted in the present paper is
represented by 4 two-electron coordinate basis wavefunctions. In principle,
the coordinate bath can be large in a system where couplings between
excited and ground state orbitals are not negligible. This is an
interesting question for future investigation: how will couplings
to excited orbitals affect the non-unitary spin dynamics? The other
interesting generalization of the present model is inclusion of real
environment effects, i.e., the real stochastic bath representing the interaction of
electron spins with the semiconductor medium. We will consider these and other
generalizations in a future publications.

In the \emph{pseudo}-spin model, where $\vec{B}$-field inhomogeneity results
in first-order effects, we have estimated the contribution of the spin-orbit
interaction to the effective \emph{pseudo}-spin Hamiltonian, namely the
Dzyaloshinskii-Moriya spin-orbit interaction term, and have suggested a
two-step mechanism: coupling between the singly-occupied singlet state and
triplet states occurs via intermediate, double-occupancy states (direct
coupling between these states turns out to be zero due to orthogonality of
the orbitals involved in the transition). Our calculations predict a smaller
magnitude of the spin-orbit interaction as compared to the estimates
of Ref.~\onlinecite{Kav01}, but are consistent with the results of
Ref.~\onlinecite{GK03}.

In our second application we demonstrated, in Figs.~\ref{Fig11} and \ref{Fig12}, 
non-unitary effects due to $\Delta \vec{B}\neq 0$ in a system of electrons
trapped above a liquid helium surface, namely the spin-based quantum
computing proposal by Lyon \cite{Lyon06}. A more thorough
investigation of spin dynamics in this system is left for a future publication.

In conclusion, we note that the two-electron spin-density matrix description
advocated in this paper is expected to be useful when electrons trapped in
QDs are not spatially resolved or resolvable. The spin-dynamics is then
completely described by the spin-density matrix $\rho $. Although the $\rho $%
-dynamics becomes non-unitary in general (at $\Delta \vec{B}\neq 0$), it is
controllable by modulating the interaction parameters, $J_{H}$, $\vec{B}%
_{av} $, $\Delta \vec{B}$. Since the non-unitarity comes from the magnetic
field inhomogeneities, $\Delta \vec{B}$ and/or $\delta \vec{B}_{so}$, and
since the $\rho $-dynamics is quadratically protected from these fields,
this might prove to be important in practical quantum computing as
minimizing coupling between spin and orbital degrees of freedom \emph{%
quadratically} improves the fidelity of unitary gates in the $\rho $-state
space.

\begin{acknowledgments}
This work was supported by NSF Grant No.\ CCF-0523675.
\end{acknowledgments}


\begin{thebibliography}{10}

\bibitem{Loss:98}
{D. Loss and D.P. DiVincenzo}, Phys. Rev. A {\bf 57},  120  (1998).

\bibitem{Kane:98}
{B.E. Kane}, Nature {\bf 393},  133  (1998).

\bibitem{Vrijen:00}
{R. Vrijen, E. Yablonovitch, K. Wang, H.W. Jiang, A. Balandin, V. Roychowdhury,
  T. Mor, and D. DiVincenzo}, Phys. Rev. A {\bf 62},  012306  (2000).

\bibitem{HuSarma01}
{X. Hu and S. Das Sarma}, Phys. Rev. A {\bf 61},  062301  (2000).

\bibitem{Schliemann01}
{J. Schliemann, D. Loss, and A.H. MacDonald}, Phys. Rev. B {\bf 63},  085311
  (2001).

\bibitem{HuSarma02}
{X. Hu and S. Das Sarma}, Phys. Rev. A {\bf 66},  012312  (2002).

\bibitem{Koiller02}
{B. Koiller, X. Hu, and S. Das Sarma}, Phys. Rev. B {\bf 66},  115201  (2002).

\bibitem{Kaplan04}
{T.A. Kaplan and C. Piermarocchi}, Phys. Rev. B {\bf 70},  161311(R)  (2004).

\bibitem{Scarola}
{V.W. Scarola and S. Das Sarma}, Phys. Rev. A {\bf 71},  032340  (2005).

\bibitem{He05}
{L. He, G. Bester, and A. Zunger}, eprint arXiv:cond-mat/0503492.

\bibitem{Hu}
{X. Hu},  in {\em {Quantum Coherence: From Quarks to Solids}}, Vol.~689 of {\em
  {Lecture Notes in Physics}} ({Springer}, {Berlin}, 2006), p.\ 83.

\bibitem{KSS97}
{J.M. Kikkawa, I.P. Smorchkowa, N. Samarth, and D.D. Awschalom}, Science {\bf
  277},  1284  (1997).

\bibitem{KA98}
{J.M. Kikkawa and D.D. Awschalom}, Phys. Rev. Lett. {\bf 80},  4313  (1998).

\bibitem{GAP98}
{J.A. Gupta, D.D. Awschalom, X. Peng, and A.P. Alivasatos}, Phys. Rev. B {\bf
  59},  R10421  (1998).

\bibitem{AS02}
{D.D. Awschalom and N. Samarth},  in {\em {Semiconductor Spintronics and
  Quantum Computation}}, edited by {D.D. Awschalom, D. Loss, N. Samarth}
  ({Springer-Verlag}, Berlin, 2002), pp.\ 147--193.

\bibitem{KLI07}
{S.D. Kunikeev and D.A. Lidar},  eprint arXiv:cond-mat/0708.0644.

\bibitem{Kav01}
{K.V. Kavokin}, Phys. Rev. B {\bf 64},  075305  (2001).

\bibitem{Aue}
{A. Auerbach}, {\em {Interacting Electrons and Quantum Magnetism}}
  ({Springer-Verlag}, {New-York}, 1994).

\bibitem{Hub1}
{J. Hubbard}, Proc. Roy. Soc. London Ser. A {\bf 276},  238  (1963).

\bibitem{Hub2}
{J. Hubbard}, Proc. Roy. Soc. London Ser. A {\bf 277},  237  (1964).

\bibitem{Hub3}
{J. Hubbard}, Proc. Roy. Soc. London Ser. A {\bf 281},  401  (1964).

\bibitem{Tak}
{M. Takahashi}, J. Phys. C {\bf 10},  1289  (1977).

\bibitem{Mac}
{A.H. MacDonald, S.M. Girvin, and D. Yoshioka}, Phys. Rev. B {\bf 37},  9753
  (1988).

\bibitem{Dzy}
{I. Dzyaloshinskii}, {J. Phys. Chem. Solids} {\bf 4},  241  (1958).

\bibitem{Mor}
{T. Moriya}, Phys. Rev. {\bf 120},  91  (1960).

\bibitem{Jacak}
{L. Jacak, P. Hawrylak, and A. W\'{o}js}, {\em {Quantum Dots}}
  ({Springer-Verlag}, {Berlin}, 1998).

\bibitem{Messiah}
{A. Messiah}, {\em {Quantum Mechanics, Vol. II}} ({North-Holland Publishing
  Company}, {Amsterdam}, 1962).

\bibitem{GK03}
{L.P. Gor'kov and P.L. Krotkov}, Phys. Rev. B {\bf 67},  033203  (2003).

\bibitem{GP63}
{L.P. Gor'kov and L.P. Pitaevskii}, {Sov. Phys. Dokl.} {\bf 8},  788  (1964).

\bibitem{HF64}
{C. Herring and M. Flicker}, Phys. Rev. A {\bf 134},  362  (1964).

\bibitem{Dres55}
{G. Dresselhaus}, Phys. Rev. {\bf 100},  580  (1955).

\bibitem{BR84}
{Y.A. Bychkov and E.I. Rashba}, J. Phys. C {\bf 17},  6039  (1984).

\bibitem{AS64}
{M. Abramowitz and I.A. Stegun (Eds.)}, {\em {Handbook of Mathematical
  Functions, 3rd Ed.}} ({National Bureau of Standards, Applied Mathematical
  Series}, {USA}, 1964).

\bibitem{DP71}
{M.I. Dyakonov and V.I Perel}, {Sov. Phys. Solid State} {\bf 13},  3023
  (1972).

\bibitem{Sla63}
{J.C. Slater}, {\em {Electronic Structure of Molecules}} ({McGraw-Hill}, {New
  York}, 1963).

\bibitem{AL87}
{R. Alicki and K. Lendi}, {\em {Quantum Dynamical Semigroups and Applications,
  Lecture Notes in Physics, V. 286}} ({Springer-Verlag}, {Berlin}, 1987).

\bibitem{Bacon:99a}
{D. Bacon, J. Kempe, D.A. Lidar and K.B. Whaley}, Phys. Rev. Lett. {\bf 85},
  1758  (2000).

\bibitem{Kempe:00}
{J. Kempe, D. Bacon, D.A. Lidar, and K.B. Whaley}, Phys. Rev. A {\bf 63},
  042307  (2001).

\bibitem{LidarWu:01}
{D.A. Lidar and L.-A. Wu}, Phys. Rev. Lett. {\bf 88},  017905  (2002).

\bibitem{Cor90}
{J.F. Cornwell}, {\em {Group Theory in Physics, V. II}} ({Academic Press},
  {London}, 1990).

\bibitem{DK02}
{R.I. Dzhioev, K.V. Kavokin, V.L. Korenev, M.V. Lazarev, B.Ya. Meltser, M.N. Stepanova, B.P. Zakharchenya, D. Gammon, D.S. Katzer}, eprint arXiv:cond-mat/0208083.

\bibitem{Lyon06}
{S.A. Lyon}, Phys. Rev. A {\bf 74},  052338  (2006).

\bibitem{PD99}
{P.M. Platzman and M.I. Dykman}, Science {\bf 284},  1967  (1999).

\bibitem{DP00}
{M.I. Dykman and P.M. Platzman}, Fortschr. Phys. {\bf 48},  1095  (2000).

\bibitem{DPS03}
{M.I. Dykman, P.M. Platzman, and Seddighrad}, Phys. Rev. B {\bf 67},  155402
  (2003).

\bibitem{DPS04}
{M.I. Dykman, P.M. Platzman, and Seddighrad}, Physica E {\bf 22},  767  (2004).

\end{thebibliography}

\end{document}